\documentclass[11pt]{article}
\usepackage{amsmath,bm,bbm}
\usepackage{geometry}

\def\gtwid{\mathrel{\raise.3ex\hbox{$>$\kern-.75em\lower1ex\hbox{$\sim
$}}}}
\def\vio{\mathrel{\hbox{$R$\kern-.60em\hbox{$/
$}}}}
\usepackage{graphicx}
\usepackage{amssymb}
\def\lsim{\mathrel{\raise.3ex\hbox{$<$\kern-.75em\lower1ex\hbox{$\sim$}}}}
\def\gsim{\mathrel{\raise.3ex\hbox{$>$\kern-.75em\lower1ex\hbox{$\sim$}}}}

\newcommand{\be}{\begin{equation}}
\newcommand{\ee}{\end{equation}}
\newcommand{\bea}{\begin{eqnarray}}
\newcommand{\eea}{\end{eqnarray}}

\begin{document}

\title
{Multi-parameter approach to R-parity violating SUSY couplings}
\author{ E. M. Sessolo$^{1}$, F. Tahir$^{2}$, D. W. McKay$^{1}$\\[2ex]
\small\it ${}^{1}$Department of Physics and Astronomy, University of
Kansas, Lawrence, KS 66045, U.S.A\\
\small\it ${}^{2}$Department of Physics, Comsats Institute of Information Technology, Islamabad 45320, Pakistan}

\date{}

\maketitle

\begin{abstract}
We introduce and implement a new, extended approach to placing bounds on trilinear R-parity
violating ($\vio{}$) couplings. We focus on a limited set of leptonic and semi-leptonic processes involving neutrinos, combining multidimensional plotting and cross-checking constraints from different experiments. This allows us to explore new regions of parameter space and to relax a number of bounds given in the literature. We look for qualitatively different results compared to those obtained previously using the assumption that a single coupling dominates the R-parity violating contributions to a process (SCD). By combining results from several experiments, we identify regions in parameter space where two or more parameters approach their maximally allowed values. In the same vein, we show a circumstance where consistency between independent bounds on the same combinations of trilinear coupling parameters implies mass constraints among slepton or squark masses. 
Though our new bounds are in most cases weaker than the SCD bounds, the largest deviations we find on individual parameters are factors of two, thus indicating that a conservative, order of magnitude bound on an individual coupling is reliably estimated by making the SCD assumption.

\end{abstract}

\section{Introduction}

It is well known that the standard model (SM) admits some
``accidental'' symmetries such as the separate conservation of
baryon ($B$) and lepton ($L$) number. In other words, the
requirement of gauge-invariance and renormalizability of the
operators that appear in the Lagrangian does not allow the
presence of terms that violate baryon or lepton number
conservation. In the framework of the Minimal Supersymmetric
Standard Model (MSSM) this is no longer true. In this model, operators that carry the same baryon and lepton number of the
standard model, but different spin or mass dimension (the
superpartners), violate $B$ and $L$ conservation, which can be enforced by hand with the introduction of R-parity.
This additional discrete symmetry of the spinorial charges allows the lightest supersymmetric particle (LSP) to remain stable and is defined as \cite{Farrar:1978xj}:

\begin{equation}
    R=(-1)^{3(B-L)-2S},\label{R-parity}
\end{equation}
with $S$ being the spin quantum number.
All the standard model particles have $R=1$, while their superpartners have $R=-1$.

The phenomenological signatures of an unstable LSP have been investigated extensively in a variety of papers, both at lepton \cite{Dimopoulos:1988jw,Barger:1989rk} and hadron \cite{Dreiner:1991pe} colliders. Generally the signatures are the consequences of new interaction terms that arise in the superpotential or in the soft supersymmetry (SUSY) breaking part of the Lagrangian when the assumption of R-parity is lifted. Wide attention has been given to extracting bounds on these new couplings from precision tests of the standard model and from cosmological constraints. The extent of the literature on the subject is daunting: we refer the reader to Ref. \cite{Barbier:2004ez} and references therein for a comprehensive review. 

Given the impact that a precise determination of the coupling size has on the phenomenological consequences, we think it is important to obtain them in the greatest possible generality. In this paper, we do so by relaxing some of the assumptions that are commonly used in the literature. After reviewing the form of the R-breaking couplings and deriving the effective Lagrangians of interest, in Sec. 2 we describe the assumptions commonly used in the literature and introduce our extended approach. In Sec. 3 we derive new bounds on the R-breaking couplings from leptonic processes, while the bounds from semi-leptonic processes are treated in Sec 4. In Sec. 5 we summarize our results and conclusions. We use new PDG2008 \cite{Amsler:2008zz} data to obtain bounds at 2$\sigma$ that are, in cases where new data has become available, more stringent than the existing ones under the standard SCD assumptions.

\section{R-parity violating couplings and low energy effective Lagrangians}

There is no theoretical argument that
prevents the superpotential from having the following bilinear or trilinear
terms\footnote{Here and throughout we follow the conventions and notations of Ref. \cite{Baer:2006rs}.}:
\begin{equation}
    \hat{f}_{T}=\lambda_{ijk}\epsilon_{ab}\hat{L}^{a}_{i}\hat{L}^{b}_{j}\hat{E}^{C}_{k}+
        \lambda'_{ijk}\epsilon_{ab}\hat{L}^{a}_{i}\hat{Q}^{b}_{j}\hat{D}^{C}_{k}+
            \lambda''_{ijk}\epsilon_{lmn}\hat{U}^{Cl}_{i}\hat{D}^{Cm}_{j}\hat{D}^{Cn}_{k}\label{f_T}
\end{equation}
and
\begin{equation}
    \hat{f}_{B}=\mu'_{i}\epsilon_{ab}\hat{L}^{a}_{i}\hat{H}^{b}_{u},\label{f_B}
\end{equation}
where the carets label the superfields corresponding to the
standard model fields, the indices $i,j,k=1,2,3$ label the
fermionic generations, $a,b=1,2$ are $SU(2)$-doublet indices,
while $l,m,n=1,2,3$ are $SU(3)$-triplet indices. The
$\lambda_{ijk}$ couplings are antisymmetric in $i,j$ due to the antisymmetry in $a,b$, imposed by $SU(2)$, while the $\lambda''_{ijk}$
are antisymmetric in $j,k$ due to the complete antisymmetry of
$\epsilon_{lmn}$, required by $SU(3)$. One can see that the first and second terms in
Eq. (\ref{f_T}), and Eq. (\ref{f_B}) violate $L$ conservation,
while the third term in Eq. (\ref{f_T}) violates $B$ conservation.
On the other hand, phenomenological considerations show that the
trilinear terms in $\lambda_{ijk}$ and $\lambda''_{ijk}$ cannot be
simultaneously present with values large enough to affect the processes we study here, otherwise squark-exchange would
lead to unacceptable rates for proton decay \cite{Dimopoulos:1981dw,Bhattacharyya:1998bx}.

Eqs. (\ref{f_T}) and (\ref{f_B}) introduce 48 new complex
parameters to the MSSM: 3 dimensionful parameters from the
bilinear couplings, and $9+27+9=45$ dimensionless parameters from
the trilinear. Along with the
superpotential terms, $B$ and $L$ can also be violated by
51 additional soft SUSY-breaking terms in the Lagrangian. Since they are not pertinent to the following discussion we will not write them explicitly here. They can be found in Ref. \cite{Barbier:2004ez}, along with a discussion of the choice of bases in which the bilinear term in the $\vio{}$ superpotential, Eq.
(\ref{f_B}), is rotated away by an $SU(4)$ transformation, so that the sneutrinos acquire a vacuum expectation value under electroweak symmetry breaking \cite{hall&suzuki,Lee:1984kr,Dawson:1985vr}.




Consistent with the existing literature on trilinear $\vio{}$ bounds as reviewed in \cite{Barbier:2004ez}, here we choose to work in the mass basis, assume all bilinear $\vio{}$ terms in the tree-level Lagrangian are absent and base our analysis solely on the trilinear terms.

Since R-parity violating terms are neither forbidden by gauge
invariance nor by renormalizability, but rather depend on
phenomenological consistency, one can wonder to what extent
R-parity could be broken, i.e. how big are the
couplings appearing in Eqs. (\ref{f_T}) and (\ref{f_B}). Restricting our discussion to Eq. (\ref{f_T}), determinations of the couplings' size are generally obtained
in the literature by comparing an effective Lagrangian expressed
in terms of the $\lambda$, $\lambda'$ and $\lambda''$ couplings
with the neutral and charged current interaction effective Lagrangian that describes fundamental tests of the standard model. 
We largely confine ourselves in this paper to flavor-conserving cases, to keep the presentation focused.
The most general effective Lagrangian for fermion-fermion neutral current
interactions $\bar{l}l\rightarrow\bar{f}f$ at low energies reads:
\begin{equation}
    \mathcal{L}=-4\sqrt{2}G_{F}\bar{l}\gamma^{\mu}[(g^{l}_{L}+\epsilon^{l}_{L})L+(g^{l}_{R}+\epsilon^{l}_{R})R]l
    \bar{f}\gamma_{\mu}[(g^{f}_{L}+\epsilon^{f}_{L})L+(g^{f}_{R}+\epsilon^{f}_{R})R]f,\label{flav-cons_Lagr}
\end{equation}
where $G_{F}=\sqrt{2}g^{2}/(8M_{W}^{2})$ is the Fermi coupling constant, $L=(1-\gamma_{5})/2$ and $R=(1+\gamma_{5})/2$ are the chiral projectors,
$g_{L}$ and $g_{R}$ are the
coupling to the chiral components of the fundamental spinors, and
the $\epsilon$'s describe the ``non-standard'' part of the
interactions. One requires that the R-breaking
contributions do not exceed the limit imposed by the precision of
the experimental measurements, thus obtaining bounds on the
couplings.

As we have mentioned above, the simultaneous presence of leptonic and hadronic R-parity violating couplings is tightly constrained experimentally by the stability of the proton. One may then choose to consider either $\lambda_{ijk}$ couplings, or $\lambda''_{ijk}$ couplings to be negligible.
In this paper we deal strictly with processes that involve $\lambda$ and $\lambda'$, as their corresponding experimental signatures are clearer and the reported uncertainties are smaller.

An effective four-fermion Lagrangian, applicable to processes at energies small compared to the weak scale, can be obtained from the superpotential of Eq. (\ref{f_T}):
\begin{equation}
    \mathcal{L}\ni-\frac{1}{2}\sum_{r,s}
    \left[\left(\frac{\partial^{2}\hat{f}}{\partial\hat{\mathcal{S}}_{r}\partial\hat{\mathcal{S}}_{s}}\right)_{\hat{\mathcal{S}}=\mathcal{S}}
    \bar{\psi}_{r}L\psi_{s}+
    \left(\frac{\partial^{2}\hat{f}}{\partial\hat{\mathcal{S}}_{r}\partial\hat{\mathcal{S}}_{s}}\right)^{\dag}_{\hat{\mathcal{S}}=\mathcal{S}}
    \bar{\psi}_{r}R\psi_{s}\right],\label{deriv_lagr}
\end{equation}
where $r$, $s$ span the superfields $\hat{\mathcal{S}}$ of the superpotential, and $\psi_{r,s}$ are the Majorana fermion fields entering the supermultiplets. The part involving semi-leptonic interactions
is given by the second term in Eq. (\ref{f_T}):
\begin{equation}
    \hat{f}\ni\lambda'_{ijk}(\hat{\nu}_{i}\hat{d}_{j}\hat{D}^{C}_{k}-\hat{e}_{i}\hat{u}_{j}\hat{D}^{C}_{k}).\label{selep-tri}
\end{equation}
Application of Eq. (\ref{deriv_lagr}) to this term yields
\begin{eqnarray}
    \mathcal{L}&\ni&-\lambda'_{ijk}(\tilde{d}^{\dag}_{Rk}\bar{\psi}_{\nu_{i}}L\psi_{d_{j}}-\tilde{d}^{\dag}_{Rk}\bar{\psi}_{e_{i}}L\psi_{u_{j}}+
    \tilde{\nu}_{i}\bar{\psi}_{d_{j}}L\psi_{D^{C}_{k}}-\tilde{e}_{Li}\bar{\psi}_{u_{j}}L\psi_{D^{C}_{k}}\nonumber\\
     &
     &+\tilde{d}_{Lj}\bar{\psi}_{\nu_{i}}L\psi_{D^{C}_{k}}-\tilde{u}_{Lj}\bar{\psi}_{e_{i}}L\psi_{D^{C}_{k}})+\textrm{ h.c.}\label{Majorana_Lagr}
\end{eqnarray}
The vertices can be obtained by defining Dirac spinors as
\begin{eqnarray}
    d (u,e)\equiv L\psi_{d (u,e)}+R\psi_{D^{C}(U^{C},E^{C})}& &d^{C}(u^{C},e^{C})\equiv
    R\psi_{d (u,e)}+L\psi_{D^{C}(U^{C},E^{C})},\label{Dirac_eud}\\
    \nu\equiv L\psi_{\nu}& &\nu^{C}\equiv
    R\psi_{\nu},\label{Dirac_nu}
\end{eqnarray}
so that one gets for the interaction part of the Lagrangian,
\begin{eqnarray}
    \mathcal{L}_{\lambda}&=&-\lambda'_{ijk}[\tilde{d}^{\dag}_{Rk}\bar{\nu^{C}_{i}}Ld_{j}+\tilde{d}_{Lj}\bar{d}_{k}L\nu_{i}+
    \tilde{\nu}_{i}\bar{d_{k}}Ld_{j}-\tilde{d}^{\dag}_{Rk}\bar{e^{C}_{i}}Lu_{j}\nonumber\\
     & &-\tilde{e}_{Li}\bar{d_{k}}Lu_{j}-\tilde{u}_{Lj}\bar{d_{k}}Le_{i}]-
     \lambda'^{\ast}_{ijk}[\tilde{d}_{Rk}\bar{d_{j}}R\nu^{C}_{i}+\tilde{d}^{\dag}_{Lj}\bar{\nu_{i}}Rd_{k}\nonumber\\
      & &+\tilde{\nu}^{\dag}_{i}\bar{d_{j}}Rd_{k}-\tilde{d}_{Rk}\bar{u}_{j}Re^{C}_{i}
      -\tilde{e}^{\dag}_{Li}\bar{u}_{j}Rd_{k}-\tilde{u}^{\dag}_{Lj}\bar{e_{i}}Rd_{k}].\label{Dirac_Lagr}
\end{eqnarray}

The effective Lagrangian for scalar mediated four-fermion interactions can be obtained by combining the vertices of Eq. (\ref{Dirac_Lagr}) and applying Fierz identities to the result:
\begin{eqnarray}
    \mathcal{L}_{eff}^{sl}&=&\frac{|\lambda'_{ijk}|^{2}}{2}\left[\frac{1}{m^{2}_{\tilde{d}_{Rk}}}
    (\bar{\nu}_{i}\gamma^{\mu}L\nu_{i})(\bar{d}_{j}\gamma_{\mu}Ld_{j})+\frac{1}{m^{2}_{\tilde{d}_{Rk}}}
    (\bar{e}_{i}\gamma^{\mu}Le_{i})(\bar{u}_{j}\gamma_{\mu}Lu_{j})-\frac{1}{m^{2}_{\tilde{d}_{Rk}}}
    (\bar{\nu}_{i}\gamma^{\mu}Le_{i})(\bar{d}_{j}\gamma_{\mu}Lu_{j})\right.\nonumber\\
     & &-\frac{1}{m^{2}_{\tilde{d}_{Lj}}}
    (\bar{\nu}_{i}\gamma^{\mu}L\nu_{i})(\bar{d}_{k}\gamma_{\mu}Rd_{k})-\frac{1}{m^{2}_{\tilde{\nu}_{Li}}}(\bar{d}_{j}\gamma^{\mu}Ld_{j})(\bar{d}_{k}\gamma_{\mu}Rd_{k})
    -\frac{1}{m^{2}_{\tilde{e}_{Li}}}
    (\bar{u}_{j}\gamma^{\mu}Lu_{j})(\bar{d}_{k}\gamma_{\mu}Rd_{k})\nonumber\\
     & &\left.-\frac{1}{m^{2}_{\tilde{u}_{Lj}}}
    (\bar{e}_{i}\gamma^{\mu}Le_{i})(\bar{d}_{k}\gamma_{\mu}Rd_{k})\right].\label{eff_Lagr_sl}
\end{eqnarray}

The effective Lagrangian of Eq. (\ref{eff_Lagr_sl}) introduces 135 independent parameters: 9 combinations in any two of the indices times 3 combinations in the remaining index which runs through the families of 5 possible exchanged sparticles.
The leptonic interaction effective Lagrangian is obtained by applying the same procedure to the first term in Eq. (\ref{f_T}):
\begin{equation}
    \hat{f}\ni\lambda_{ijk}(\hat{\nu}_{i}\hat{e}_{j}\hat{E}^{C}_{k}-\hat{e}_{i}\hat{\nu}_{j}\hat{E}^{C}_{k}).\label{lep-tri}
\end{equation}
One gets \cite{Barger:1989rk}
\begin{eqnarray}
    \mathcal{L}_{eff}^{lep}&=&\frac{|\lambda_{ijk}|^{2}}{2}\left[\left(\frac{1}{m^{2}_{\tilde{e}_{Rk}}}
    (\bar{\nu}_{i}\gamma^{\mu}L\nu_{i})(\bar{e}_{j}\gamma_{\mu}Le_{j})-\frac{1}{m^{2}_{\tilde{e}_{Rk}}}
    (\bar{e}_{i}\gamma^{\mu}L\nu_{i})(\bar{\nu}_{j}\gamma_{\mu}Le_{j})\right.\right.\nonumber\\
 & &-\frac{1}{m^{2}_{\tilde{\nu}_{Li}}}(\bar{e}_{j}\gamma^{\mu}Le_{j})(\bar{e}_{k}\gamma_{\mu}Re_{k})\left.\left.-\frac{1}{m^{2}_{\tilde{e}_{Li}}}
    (\bar{\nu}_{j}\gamma^{\mu}L\nu_{j})(\bar{e}_{k}\gamma_{\mu}Re_{k})\right)+(i\leftrightarrow j)\right],\label{eff_Lagr_l}
\end{eqnarray}
where $i<j$ is understood in Eq. (\ref{eff_Lagr_l}).
The same antisymmetry in the $i$ and $j$ indices of the $\lambda$ couplings reduces the number of effective independent couplings encompassed in Eq. (\ref{eff_Lagr_l}) with respect to the semileptonic case. There are 45: 3 free parameters in a 3$\times$3 antisymmetric matrix multiply 3 possibilities for the remaining free index that carries the dependence on the $m_{\tilde{e}_{Rk}}$; plus, there are 6$\times$3 possibilities with left-handed sneutrino exchange and 18 more possibilities with left-handed selectron exchange. 

The limits in the literature are obtained under the assumption that a single coupling dominates the R-parity violating contributions to a process (SCD). This assumption
rests on the premise that some hierarchy exists between the
leptonic, semileptonic and hadronic couplings, or between
different fermionic families. Besides, the couplings often enter as sums of squares, so that one might guess the most conservative bounds follow from this hypothesis. We found that in most cases this is not so.

It is an open question whether such a hierarchy does
indeed exist. In the absence of a theoretical guide, we apply a ``multi-parameter'' approach to placing bounds, to explore new regions of parameter space. In Sections 3 and 4 we give examples of our
approach and contrast the results to those of the SCD simplification.

\subsection{Notation and conventions}

We think it is important at this point to clarify our notation, as the originality of our contribution rests in making explicit use of some properties of $\vio{}$-couplings that are often overlooked in the literature, partly because the established notation bears some elements of ambiguity.
As far as SCD is concerned, the concept was originally formulated by Dimopoulos and Hall \cite{Dimopoulos:1988jw}. As generally applied, one assumes that a single coupling (or a single product of couplings) is much larger than the others which, therefore, can be neglected when placing bounds. As is the case in most of the corrections to the SM that involve R-parity violating couplings, more than one coupling is present and often this simultaneous presence is not clear in the notation. For example, when the process at hand involves the four-fermion interactions described by Eqs. (\ref{eff_Lagr_sl}) and (\ref{eff_Lagr_l}), the initial and final states of the scattering or decay are supposed to be completely known, whereas the exchanged sparticle, whether a squark or a slepton, can be of any generation. Thus, since this flavor is unknown, one has always to sum over the families of the sparticles compatible with the relevant vertices. So, it is important to understand that a bound that reads, for example, $|\lambda_{12k}|\leq0.15(\tilde{e}_{Rk})$ can be taken to mean either
\begin{eqnarray}
 |\lambda_{12k}|\leq0.15\left(\frac{m_{\tilde{e}_{Rk}}}{\textrm{100 GeV}}\right),& &\textrm{ (strong version) }\label{notation4a}
\end{eqnarray}
for each $k=1,2,3$, or
\begin{eqnarray}
\sqrt{\sum_{k}\left(\frac{|\lambda_{12k}|}{m_{\tilde{e}_{Rk}}}\right)^{2}}\leq0.0015\textrm{ GeV}^{-1}.& &\textrm{ (weak version) }\label{notation4b}
\end{eqnarray}
As in Eqs. (\ref{notation4a}) and (\ref{notation4b}), we adopt the standard 100 GeV scaling of sfermion masses throughout. One of the ways of implementing the SCD convention consists in setting all but one $\lambda_{12k}$ in Eq. (\ref{notation4b}) to zero, thus effectively obtaining Eq. (\ref{notation4a}). The strong version produces bounds that are obviously more conservative than the weak one, so we will display the form (\ref{notation4a}) every time we place a new bound on a coupling, with the caveat that the reader can interpret it in the form of Eq. (\ref{notation4b}). We will state explicitly when we make an exception to this rule.


If the initial and final states of the four-fermion process involve the same vertices, the R-breaking couplings enter the process only through their modulus squared. In the literature it is then customary to express the corrections to the SM as functions of simplified quantities: $r_{ijk}(\tilde{l}_{i})$ (but also $r_{ijk}(\tilde{l}_{j})$, $r_{ijk}(\tilde{l}_{k})$) or  $r'_{ijk}(\tilde{f}_{i})$ (but also $r'_{ijk}(\tilde{f}_{j})$, $r'_{ijk}(\tilde{f}_{k})$). In light of what we have explained above, we want to make clear that these are symbols that stand in full for:
\begin{eqnarray}
    r_{ijk}(\tilde{l}_{i})=\sum_{i}\frac{|\lambda_{ijk}|^{2}}{4\sqrt2 G_{F}m_{\tilde{l}_{i}}^{2}}&\textrm{ and }&r'_{ijk}(\tilde{f}_{i})=\sum_{i}\frac{|\lambda'_{ijk}|^{2}}{4\sqrt2 G_{F}m_{\tilde{f}_{i}}^{2}},\label{rijk}
\end{eqnarray}
where the scaling factor $4\sqrt2 G_{F}$ comes from the general form, Eq. (\ref{flav-cons_Lagr}).
Thus, they admit a sum over the flavors of the exchanged sfermion $\tilde{f}_{i}$ (or $\tilde{f}_{j}$, $\tilde{f}_{k}$) which, depending on the case, can be a slepton ($\tilde{l}_{i}$) or a squark ($\tilde{q}_{i}$). It is also clear that the value of the mass of the exchanged sparticle is always left unknown. If, instead, the SUSY process involves different vertices, then the correction to the standard model is expressed as a function of a product of couplings, of the kind $\lambda_{ijk}\cdot\lambda_{rsk}$ (equivalently, $\lambda'_{ijk}\cdot\lambda'_{rsk}$ or $\lambda_{ijk}\cdot\lambda'_{rsk}$). In these cases too, a sum over $k$ needs to be considered. Analyses that return one product as dominant are a common extension of the SCD.

We have decided to label the fermion (sfermion) generations by a number index $i$ (or $j$ or $k$) $=1,2,3$ whenever the families are summed over, as in Eqs. (\ref{notation4b}) and (\ref{rijk}), or when one of the indices is free to take any values, as in Eq. (\ref{notation4a}). But, for clarity's sake, if the bound involves just one single particular coupling we will label the generation by name so that, for example, $\tilde{e}_{R1}\leftrightarrow\tilde{e}_{R}$, $\tilde{\nu}_{L2}\leftrightarrow\tilde{\nu}_{\mu L}$,  $\tilde{u}_{L3}\leftrightarrow\tilde{t}_{L}$, and so on.  

We have mentioned above that implementation (\ref{notation4a}) of the SCD produces bounds that are more conservative. 
In most cases, though, a physical process cannot be expressed in terms of only one combination of couplings such as (\ref{rijk}). The bounds from experiment are placed generally on a function of combinations
\begin{equation}
 F\left(r_{ijk}(\tilde{l}_{i}),r_{rst}(\tilde{l}_{s}),r'_{lmn}(\tilde{q}_{n}),...\right).\label{notation}
\end{equation}
It is a common approach in the literature to set all the $r$'s of Eq. (\ref{notation}) but one to zero, so as to place bounds on the surviving combination of couplings. Moreover, one term at a time of the combination is then assumed to dominate. It is this particular implementation of the SCD that we find excessively severe, as it reduces the dimensionality of the allowed regions of parameter space thus missing any information on the combined action of different couplings embedded in the function $F$.
In the next two sections we show that allowing the full dependence on $F$ does indeed give more information and in some cases also extends the allowed bounds on the couplings.

\section{Leptonic case}

In order to show how our approach works, we start with some classical examples in the leptonic case \cite{Barger:1989rk,Barbier:2004ez}. We begin with constraints required by universality in muon and tau decays, then take up the constraints from $\nu_{\mu} e$, $\nu_{e} e$, and $\bar{\nu}_{e} e$ elastic scattering cross section measurements.

\subsection{Muon and tau decays}
 Let us consider the two following ratios:
\begin{equation}
 R_{\tau\mu}=\frac{\Gamma(\tau^{-}\rightarrow\mu^{-}\bar{\nu}_{\mu}\nu_{\tau})} {\Gamma(\mu^{-}\rightarrow e^{-}\bar{\nu}_{e}\nu_{\mu})}\label{Rtaumu}
\end{equation}
and
\begin{equation}
 R_{\tau}=\frac{\Gamma(\tau^{-}\rightarrow e^{-}\bar{\nu}_{e}\nu_{\tau})} {\Gamma(\tau^{-}\rightarrow\mu^{-}\bar{\nu}_{\mu}\nu_{\tau})},\label{Rtau}
\end{equation}
which are sensitive to violation of lepton universality.
By comparing the tree-level effective Lagrangian of Eq. (\ref{eff_Lagr_l}) with the SM, one can derive bounds on some of the $\lambda$-couplings. The SUSY processes that contribute to the decay (\ref{Rtaumu}) are shown in Figure 1. Those for (\ref{Rtau}) can be obtained by replacing $j=2\rightarrow 3$ in Figure 1b.

\begin{figure}[h!]
 \begin{center}
  \includegraphics[width=160mm, height=60mm]{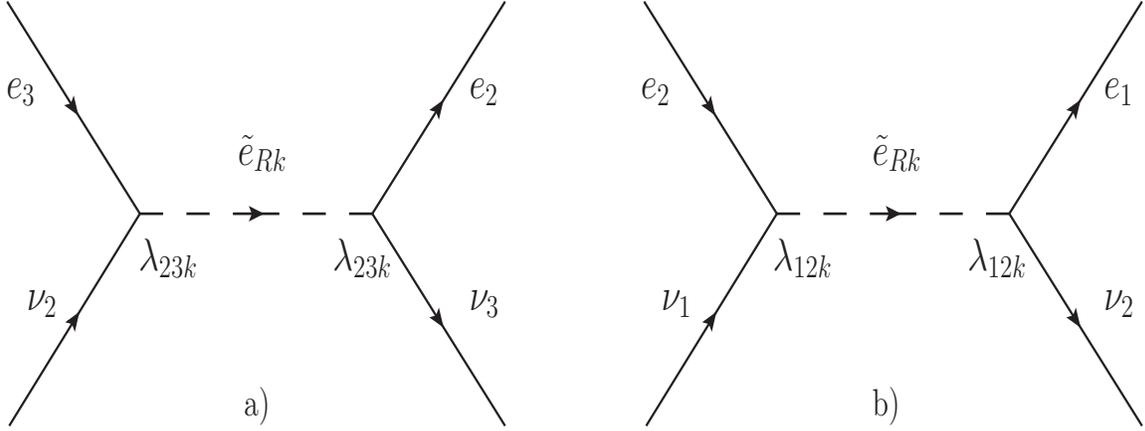}
     \caption{a) R-breaking contribution to $\tau^{-}\rightarrow\mu^{-}\bar{\nu}_{\mu}\nu_{\tau}$. b) R-breaking contribution to $\mu^{-}\rightarrow e^{-}\bar{\nu}_{e}\nu_{\mu}$.}
   \end{center}
\end{figure}

Besides, one also needs to consider the $\lambda$-dependence of the Fermi coupling constant $G_{F}$ \cite{Dimopoulos:1988jw}. As is well known, $G_{F}$ is experimentally determined from measurements of the muon lifetime. Therefore, when dealing with R-parity violating SUSY, $G_{F}$ receives a correction from the SUSY processes that contribute to $\mu$-decay, Figure 1b. The correction is given by \cite{Barger:1989rk}:
\begin{equation}
    \frac{G_{F}}{\sqrt{2}}=\frac{g^{2}}{8M^{2}_{W}}
    \left(1+\frac{M^{2}_{W}}{g^{2}m^{2}_{\tilde{e}_{Rk}}}|\lambda_{12k}|^{2}\right)\equiv\frac{g^{2}}{8M_{W}^{2}}\left[1+r_{12k}(\tilde{e}_{Rk})\right].\label{GF_corr}
\end{equation}
where a sum over repeated indices is intended, as explained in Sec. 2.1. We have also used the notation introduced in Eq. (\ref{rijk}).


Taking into account the processes of Fig. 1, Eqs. (\ref{Rtaumu}) and (\ref{Rtau}) can be expressed in terms of their SM expressions and yield, to first order in R-breaking couplings \cite{Barger:1989rk},
\begin{equation}
 R_{\tau\mu}=R_{\tau\mu}^{SM}\left\{1+2\left[r_{23k}(\tilde{e}_{Rk})-r_{12k}(\tilde{e}_{Rk})\right]\right\}\label{RtmRPV}
\end{equation}
and
\begin{equation}
 R_{\tau}=R_{\tau}^{SM}\left\{1+2\left[r_{13k}(\tilde{e}_{Rk})-r_{23k}(\tilde{e}_{Rk})\right]\right\},\label{RtRPV}
\end{equation}
again, with the conventions of Eq. (\ref{rijk}).
As explained in the discussion preceding and following Eq. (\ref{notation}), if we were to use the SCD at this point, we would consider one $r$-combination at a time and obtain a bound on each of them when the remaining couplings are put to zero. By using the measured values of $R_{\tau\mu}$ and $R_{\tau}$ \cite{Amsler:2008zz} and the standard model values after radiative corrections (\cite{Sirlin:1980nh,Marciano:1985pd} and References therein) for $R_{\tau\mu}^{SM}$ and $R_{\tau}^{SM}$, we would obtain at 2$\sigma$: $|\lambda_{23k}|\leq0.063$ $(\tilde{e}_{Rk})$ and $|\lambda_{12k}|\leq0.045$ $(\tilde{e}_{Rk})$ from $R_{\tau\mu}$, $|\lambda_{23k}|\leq0.051$ $(\tilde{e}_{Rk})$ and $|\lambda_{13k}|\leq0.048$ $(\tilde{e}_{Rk})$ from $R_{\tau}$, where the dominant uncertainty is the one on the $\tau$ lifetime, and we have used the conventions introduced in the discussion preceding Eq. (\ref{notation4a}). In principle, though reasonably well motivated and not inconsistent, \textit{there is no theoretical justification for considering one coupling at a time}, or a sum over families at a time. As Figure 2 shows, the full dependence on the couplings presents a richer structure.

\begin{figure}[h!]
 \begin{center}
  \includegraphics[width=130mm, height=60mm]{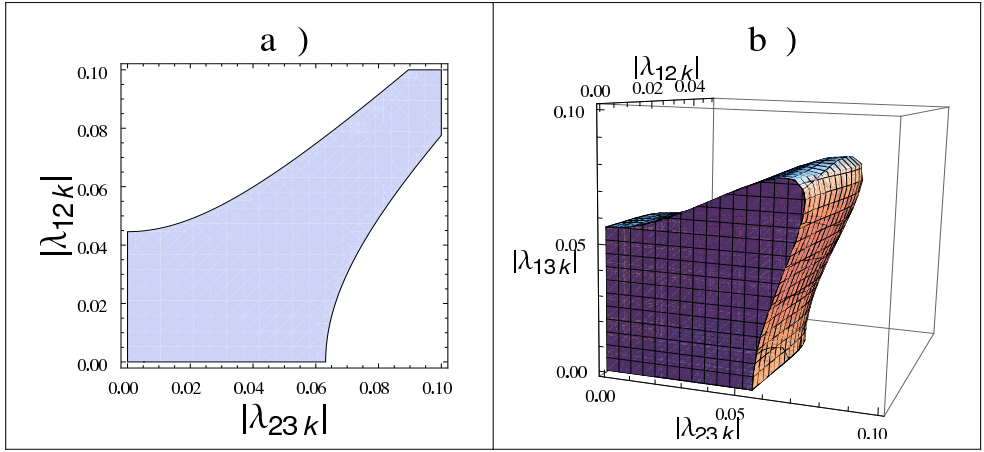}
     \caption{a) $|\lambda_{23k}|$ and $|\lambda_{12k}|$ are underconstrained at 2$\sigma$ if one takes solely into account $R_{\tau\mu}$, Eq. (\ref{RtmRPV}). b) 2$\sigma$ bound region on $\lambda_{23k}(\tilde{e}_{Rk})$, $\lambda_{12k}(\tilde{e}_{Rk})$ and $\lambda_{13k}(\tilde{e}_{Rk})$ from $R_{\tau\mu}$, $R_{\tau}$ and muon lifetime combined data. The allowed region can be enclosed in a box of size $\{|\lambda_{23k}|(\tilde{e}_{Rk}), |\lambda_{12k}|(\tilde{e}_{Rk}),|\lambda_{13k}|(\tilde{e}_{Rk})\}\leq\{0.075,0.043,0.082\}$.}
   \end{center}
\end{figure}

Figure 2a shows that Eq. (\ref{RtmRPV}) admits degeneracies on the couplings. When taken together, $|\lambda_{23k}|$ and $|\lambda_{12k}|$ can be taken arbitrarily large\footnote{Consistent with the underlying perturbation expansion.}, since they cancel each other. A similar picture holds for $R_{\tau}$, as Eq. (\ref{RtRPV}) has the same form as Eq. (\ref{RtmRPV}). Thus, an approach extended beyond SCD consists in trying to limit and reduce those degeneracies by combining different experiments that involve the same couplings.

In particular, the measurement of the muon lifetime can be used to determine a first bound on the sum of couplings $\lambda_{12k}$, when a right-handed charged slepton is exchanged, Eq. (\ref{GF_corr}). The result is dependent on radiative corrections and on the renormalization scheme. Expressions for $\lambda_{12k}/m^{2}_{\tilde{e}_{Rk}}$ can be derived in the \textit{on shell}, $\overline{MS}$ and Novikov-Okun-Vysotsky (NOV) renormalization schemes \cite{Novikov:1992rj}, and are given by
\begin{equation}
    \frac{|\lambda_{12k}|^{2}}{m^{2}_{\tilde{e}_{Rk}}}=\frac{8G_{F}}{\sqrt{2}}\left[\frac{M_{Z}^{2}\sqrt{2}G_{F}\hat{\rho}|_{\textrm{scheme}}(\sin^{2}\theta_{W}\cos^{2}\theta_{W})|_{\textrm{scheme}}(1-\Delta r|_{\textrm{scheme}})} {\pi \alpha}-1\right],\label{renorm}
\end{equation}
where the scheme dependence of $\hat{\rho}$, $\sin^{2}\theta_{W}$ and $\Delta r$ can be found in Table 1.
\begin{table}
\begin{center}
\begin{tabular}{|c||c|c|c|}
    \hline
    Scheme & $\sin^{2}\theta_{W}$ & $\Delta r$ & $\hat{\rho}$\\
    \hline \hline
    On shell (o.s.) & $1-\frac{M_{W}^2}{M_{Z}^2}$ & $1-\frac{\alpha}{\hat{\alpha}(M_{Z})}-\frac{\rho_{t}}{\tan^{2}\theta_{W}|_{\textrm{o.s.}}}$ & 1\\
    \hline
    $\overline{MS}$ & $\left(1+\frac{\rho_{t}}{\tan\theta_{W}|_{\textrm{o.s.}}}\right)
\left(\sin^{2}\theta_{W}|_{\textrm{o.s.}}\right)$ & $1-\frac{\alpha}{\hat{\alpha}(M_{Z})}+...$\cite{Marciano:1990dp} & 1.01023(22)\\
    \hline
NOV & $\frac{1}{2}-\left(\frac{1}{4}-\frac{\pi\alpha(M_{Z})}{\sqrt{2}G_{F}M_{Z}^{2}}\right)^{1/2}$ & $1-\frac{\alpha}{\alpha(M_{Z})}$ & 1\\
    \hline
\end{tabular}
\caption{Analytic expressions for $\sin^{2}\theta_{W}$, $\Delta r$ and $\hat{\rho}$ in the \textit{on shell}, $\overline{MS}$ and NOV renormalization schemes. $\rho_{t}=3G_{F}m_{t}^{2}/8\sqrt{2}\pi^{2}$. We use the following average values at 1$\sigma$ \cite{Amsler:2008zz}: $\sin^{2}\theta_{W}|_{\overline{MS}}=0.23119(14)$; $\hat{\alpha}^{-1}(M_{Z})=127.925(16)$; $\alpha^{-1}(M_{Z})=128.91(2)$;  $\sin^{2}\theta_{W}|_{\textrm{NOV}}=0.23108(5)$; $\Delta r|_{\textrm{o.s.}}=0.0369(14)$; $\Delta r|_{\overline{MS}}=0.06962(12)$. The ellipsis indicates non-leading order terms that can be found in \cite{Marciano:1990dp}.}
\end{center}
\end{table}
According to the different renormalization scheme, we find the 1$\sigma$-bounds on $\lambda_{12k}$ obtained from the muon lifetime:
\begin{eqnarray}
 \textit{(on shell)}& &|\lambda_{12k}|\textrm{ is excluded}\label{lam12kos}\\
 (\overline{MS})& &|\lambda_{12k}|\leq0.029\left(\frac{m_{\tilde{e}_{Rk}}}{\textrm{100 GeV}}\right)\label{lam12kMSbar}\\
 \textrm{(NOV)}& &|\lambda_{12k}|\leq0.011\left(\frac{m_{\tilde{e}_{Rk}}}{\textrm{100 GeV}}\right).\label{lam12kNOV}
\end{eqnarray}
And at 2$\sigma$:
\begin{eqnarray}
 \textit{(on shell)}& &|\lambda_{12k}|\leq0.031\left(\frac{m_{\tilde{e}_{Rk}}}{\textrm{100 GeV}}\right)\label{lam12kos2}\\
 (\overline{MS})& &|\lambda_{12k}|\leq0.037\left(\frac{m_{\tilde{e}_{Rk}}}{\textrm{100 GeV}}\right)\label{lam12kMSbar2}\\
 \textrm{(NOV)}& &|\lambda_{12k}|\leq0.015\left(\frac{m_{\tilde{e}_{Rk}}}{\textrm{100 GeV}}\right).\label{lam12kNOV2}
\end{eqnarray}
We can therefore decide to use this bound to limit the degeneracies present in $R_{\tau\mu}$ and $R_{\tau}$. Figure 2b shows the 2$\sigma$-allowed region of parameter space in $\lambda_{23k}$, $\lambda_{12k}$ and $\lambda_{13k}$ when the PDG2008 data for $R_{\tau\mu}$, $R_{\tau}$ and muon lifetime are combined. We use the $\overline{MS}$ bound, Eq. (\ref{lam12kMSbar2}), on $\lambda_{12k}$. One can see that the $\lambda$-parameters undergo a extension up to a factor of two with respect to the value obtained using the SCD. When scaled to the masses of the exchanged sleptons the 2$\sigma$ region shown in Fig. 2b can be enclosed in a box of size $\{|\lambda_{23k}|(\tilde{e}_{Rk}), |\lambda_{12k}|(\tilde{e}_{Rk}),|\lambda_{13k}|(\tilde{e}_{Rk})\}\leq\{0.075,0.043,0.082\}$. The 2$\sigma$ bounds are: 
\begin{equation}
 |\lambda_{23k}|\leq0.066\left(\frac{m_{\tilde{e}_{Rk}}}{\textrm{100 GeV}}\right)\label{l23kLep}
\end{equation}
and
\begin{equation}
 |\lambda_{13k}|\leq0.071\left(\frac{m_{\tilde{e}_{Rk}}}{\textrm{100 GeV}}\right).\label{l13kLepMS}
\end{equation}
The other schemes yield similar values.
Note that the new bound on $\lambda_{13k}$ does not include $\lambda_{133}$, since this coupling is separately and more severely bounded by the $\nu_{e}$ mass \cite{Dimopoulos:1988jw}. Similarly, there exist strong bounds on many pair-wise products of the couplings above, coming from experimental bounds on decays disallowed in the SM \cite{Barbier:2004ez,dko}. However, there are always combinations of $\lambda_{12k}$, $\lambda_{13k}$ and $\lambda_{23k}$ that are still unconstrained by the bounds on products. This comment applies to all the cases that we are considering. To list the detailed conditions takes us beyond the aim of this paper, so we leave them as implicit.

\subsection{Neutrino - electron scattering}

We now turn to the flavor diagonal neutrino - electron scattering processes $\nu_{\mu}+e \rightarrow \nu_{\mu}+e$ and $ \nu_{e}+e \rightarrow \nu_{e}+e$.  In the $\nu_{\mu}e$ and $\nu_{e}e$ examples, the energy is always large enough to neglect the electron mass in the kinematics, while in the $\bar{\nu}_{e}e$ case, the neutrino energies are in the MeV range, which requires us to keep the electron mass effects in the kinematics. The individual left -and right-handed couplings or, equivalently, axial and vector couplings, have been extracted individually in the experiments on the $\nu_{\mu}e$ case \cite{Vilain:1994qy}, making the analysis of bounds on R-parity violating parameters quite straightforward; we begin with this process.

\subsubsection{$\nu_{\mu}+e \rightarrow \nu_{\mu}+e$}
\space
Neglecting the terms proportional to the electron mass, the total cross sections for $\nu_{\mu}+e \rightarrow \nu_{\mu}+e$ and $\overline{\nu}_{\mu}+e \rightarrow \overline{\nu}_{\mu}+e$ can be written as 
\begin{equation}
\sigma (\nu_{\mu} e) = \frac{G^{2}_{F} s}{\pi} \left(g_{L}^2 + \frac{1}{3} g_{R}^2\right)
\end{equation}
and 
\begin{equation}
\sigma (\overline{\nu}_{\mu} e) = \frac{G^{2}_{F} s}{\pi} \left(g_{R}^2 + \frac{1}{3} g_{L}^2\right).
\end{equation}
The direct-channel Mandelstam variable $s = 2m_eE_{\nu}$ in the target-electron rest frame.
We can write $g_{L}$ and $g_{R}$ in terms of the weak angle and the R-parity violating parameters (Figure 3) as
\begin{equation}
g_{L}  = g_{L}^{SM} -(1+g_{L}^{SM})r_{12k}(\tilde{e}_{Rk}) 
\end{equation}
and
\begin{equation}
g_{R}  =  g_{R}^{SM}+r_{121}(\tilde{e}_{L})+r_{231}(\tilde{\tau}_{L}) - g_{R}^{SM}r_{12k}(\tilde{e}_{Rk}),
\end{equation}
where $x_{W}\equiv\sin^{2}\theta_{W}$, and $g_{L}^{SM} = x_{W}-\frac{1}{2}$ and $g_{R}^{SM}=x_{W}$ are the SM expressions for the $L$ and $R$ couplings.

\begin{figure}[h!]
 \begin{center}
  \includegraphics[width=160mm, height=60mm]{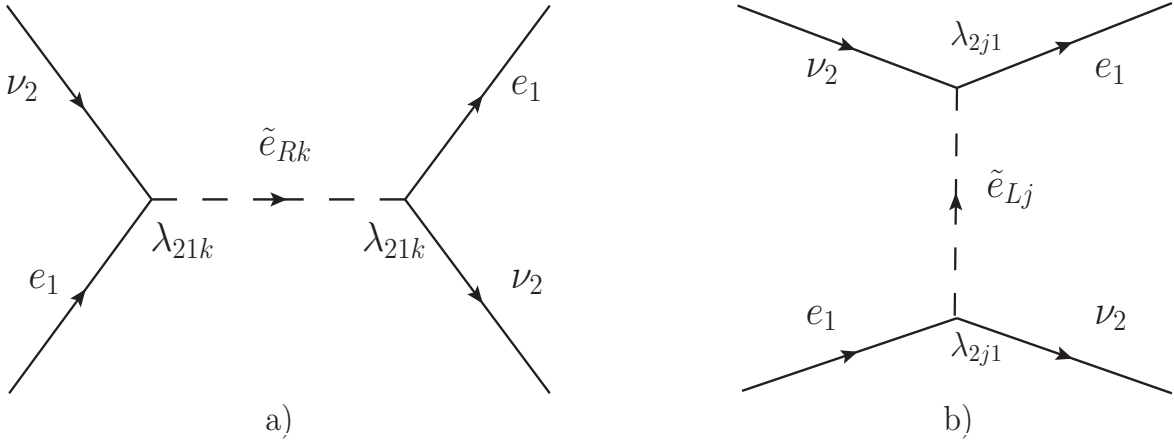}
     \caption{a) SUSY R-breaking process contributing to $g_{L}$ and $g_{R}$. b) SUSY process contributing to $g_{R}$. Here gauge invariance requires $j=1,3$.}
   \end{center}
\end{figure}

Since the experimental averages, with errors, are reported by the Particle Data Group \cite{Amsler:2008zz} for $g_{A} = g_{L} - g_{R}$ and $g_{V} = g_{L} + g_{R}$, we use these forms to obtain the bounds on the R-parity violating couplings:
\begin{equation}
g_{A}  =  g_{A}^{SM}\left[1-r_{12k}(\tilde{e}_{Rk})\right] - r_{121}(\tilde{e}_{L} )- r_{231}(\tilde{\tau}_{L})-r_{12k}(\tilde{e}_{Rk})=-0.507 \pm 0.014
\end{equation}
and
\begin{equation}
g_{V}  =  g_{V}^{SM}\left[1-r_{12k}(\tilde{e}_{Rk})\right]+r_{121}(\tilde{e}_{L})+r_{231}(\tilde{\tau}_{L})-r_{12k}(\tilde{e}_{Rk})=-0.040 \pm 0.015,
\end{equation} 
where $g_{V}^{SM}=-0.0397 \pm 0.0003$ and $g_{A}^{SM}=-0.5064 \pm 0.0001$.  The values quoted for the $g$-parameters are taken from the 2008 Particle Data Group, who point out that the CHARM II results \cite{Vilain:1994qy} dominate the average values.

Including $\lambda_{12k}(\tilde{e}_{Rk})$, Eq. (\ref{lam12kMSbar2}), in the 2$\sigma$ joint bounds, we find the corresponding upper bound to be:
\begin{equation}
 \sqrt{\left[|\lambda_{121}|\left(\frac{\textrm{ 100 GeV}}{m_{\tilde{e}_{L}}}\right)\right]^2+\left[|\lambda_{231}|\left(\frac{\textrm{ 100 GeV}}{m_{\tilde{\tau}_{L}}}\right)\right]^2} \leq 0.130. \label{numu-e-bnd}
\end{equation}
To put Eq. (\ref{numu-e-bnd}), a bound on the sum of squares of couplings divided by scaled masses, in the context of other bounds, we can use the bound from \cite{dko}, updated to 2008 data \cite{Amsler:2008zz}:
\begin{equation}
\left|\lambda_{121}\left(\frac{\textrm{ 100 GeV}}{m_{\tilde{\nu}_{\mu L}}}\right)\times \lambda_{231}\left(\frac{\textrm{ 100 GeV}}{m_{\tilde{\nu}_{\mu L}}}\right)\right| \leq 3.0 \times 10^{-4}. \label{prod-bnd}
\end{equation}
The bound in Eq. (\ref{prod-bnd}) combines the experimental bound on the decay rate for $\tau \rightarrow ee\overline{e}$ with its representation in the ``double coupling dominance convention'' for R-parity violating trilinear couplings \cite{dko}.  The representation of the decay involves the sum of squares of five coupling products, and the convention, in this case, serves to place the weakest bound on each product by assuming all the others are effectively zero. This example, though not in line with our restriction to flavor-conserving processes, allows us to discuss the implications for sfermion masses that follow from $\vio{}$ bounds.
Because three unknown masses appear in Eqs. (\ref{numu-e-bnd}) and (\ref{prod-bnd}), what one can say about the implications of the bounds for the $\lambda$ parameters is limited, even if the individual $\vio{}$ couplings entering the two equations are the same. Concisely put, one can say that whenever the sneutrino mass satifies
\begin{equation}
 m_{\tilde{\nu}_{\mu L}}\geq \frac{0.130}{\sqrt{2\times 3.0\times 10^{-4}}} \sqrt{m_{\tilde{e}_{L}}\times m_{\tilde{\tau}_{L}}}\approx 5.3\sqrt{m_{\tilde{e}_{L}}\times m_{\tilde{\tau}_{L}}}\label{sneu-ineq} 
\end{equation}
the bound of Eq. (\ref{numu-e-bnd}) is more restrictive than that of Eq. (\ref{prod-bnd}), which becomes irrelevant.
If, instead, Eq. (\ref{sneu-ineq}) is not satisfied, the above bounds have to be considered together, because the hyperbola described by Eq. (\ref{prod-bnd}) will cut through the elliptical region defined by Eq. (\ref{numu-e-bnd}), and part of the region allowed by Eq. (\ref{numu-e-bnd}) will be prohibited by Eq. (\ref{prod-bnd}). 

Unless we invoke some theoretical prejudice about the relative mass scales, we cannot conclude more than that. Only if one of the inequalities includes a lower bound, does the combination of bounds lead to a general condition on the masses.  We will see an illustration of this situation below, when considering the combined bounds on $\nu_e e$ and $\overline{\nu}_e e$ scattering at 1$\sigma$.

\subsubsection{$\nu_{e}+e \rightarrow \nu_{e}+e$}

Turning to the implications of data on the scattering processes $\nu_{e}+e \rightarrow \nu_{e}+e$ and $\overline{\nu}_{e}+e \rightarrow \overline{\nu}_{e}+e$ \cite{irvine,lsnd, bugey}, we must consider both high energy data, $E_{\nu} \gg m_{e}$, and low energy data, $E_{\nu} \sim m_{e}$. General, model independent analyses of bounds on non-standard interactions from these and related neutrino and electron data have recently been carried out for both non-universal and flavor-changing new physics interactions \cite{bmmv, bmmv2, bmptv}.  We focus here on the bounds on $\vio{}$ trilinear coupling parameters provided by flavor diagonal elastic $\nu_e e$ accelerator data at tens of MeV \cite{lsnd} and elastic $\overline{\nu}_e e$ reactor data at several MeV \cite{irvine}.

The LSND Collaboration provides a measurement of the total cross section for elastic scattering of the electron neutrinos off electrons.  Assuming that the final state neutrinos are also electron-type, we can use their reported value and the general expression for left handed neutrinos scattering of unpolarized electrons to set a limit. 
The general expression for a $(V-A)\bigotimes [g_L(V-A)\bigoplus g_R(V+A)]$ four-fermion interaction differential cross section reads 
\begin{equation}
 \frac{d\sigma}{dT}=\frac{2G_{F}^{2}m_{e}}{\pi}\left[g_{L}^{2}+g_{R}^{2}\left(1-\frac{T}{E_{\nu}}\right)^{2}-
g_{L}g_{R}\frac{m_{e}T}{E_{\nu}^{2}}\right],\label{diff_crx}
\end{equation}
and for the total cross section
\begin{eqnarray}
\sigma_{\nu e}&=&\frac{2 m_e E_{\nu} G^{2}_F}{\pi}\left(g_L^2 +\frac{1}{3}g_R^2 -\frac{1}{2} \frac{m_e}{E_{\nu}} g_L g_R\right)\nonumber \\
                       &=&(10.1 \pm 1.5) \times 10^{-42} E_{\nu} (\textrm{GeV}) \textrm{ cm}^2, \label{nue_bnd}
\end{eqnarray}
where $E_{\nu}$ is the neutrino energy in the rest frame of the target electron, and $T$ is the kinetic energy of the recoil electron\footnote{In view of the low precision of the experimental uncertainties, we have not included radiative corrections. For a study discussing radiative corrections and future possibilities for precision measurements, see \cite{Marciano:2003eq}.}.  The $\bar{\nu}_{e}e$ cross sections follow by interchanging $g_{L}$ and $g_{R}$ in Eqs. (\ref{diff_crx}) and (\ref{nue_bnd}). When $E_{\nu} \gg m_e$, as in the case of the LSND experiment, $m_e/E_{\nu}$ is ignorable, and the expression for the cross section simplifies to the familiar high energy form.
Including the $\vio{}$ trilinear parameters in the expressions for the $g_{L}$ and $g_{R}$ coupling coefficients, we find
\begin{eqnarray}
g_L ^{\nu_e e}&=&\left(\frac{1}{2}+x_W\right) \left[1-r_{12k}(\tilde{e}_{Rk})\right]\\
g_R^{\nu_e e}&=&x_W \left[1-r_{12k}(\tilde{e}_{Rk}) \right]+r_{121}(\tilde{\mu}_L)+r_{131}(\tilde{\tau}_L),\label{gLgR}
\end{eqnarray}
where we have considered a SUSY process like the one depicted in Figure 3b, in which $\nu_{2}\equiv\nu_{\mu}$ has to be replaced by $\nu_{1}\equiv\nu_{e}$, $\lambda_{2j1}\rightarrow\lambda_{1j1}$, and $j=2,3$, while the $\lambda_{12k}$-dependence is given by the correction to $G_{F}$, Eq. (\ref{GF_corr}).
In our study of the bound on $r_{12k}(\tilde{e}_{Rk})$ that follows from the precision measurement of muon decay and the renormalized expression for the muon decay formula, we found the bounds of Eqs. (\ref{lam12kMSbar}) and (\ref{lam12kMSbar2}) respectively at the 1$\sigma$ and 2$\sigma$ C.L. The corresponding values of $r_{12k}(\tilde{e}_{Rk})$ are so small that it can be dropped from further discussion.  The coupling coefficient $g_L$ then has its SM value, and $g_R$ is modified from the SM value by the terms that depend on $\lambda_{121}$ and $\lambda_{131}$. 
Referring to Eq. (\ref{nue_bnd}) and (\ref{gLgR}), we find the bound on the region of trilinear couplings we are after:
\begin{equation}
\sqrt{\left(|\lambda_{121}|\frac{\textrm{ 100 GeV}}{m_{\tilde{\mu}_L}}\right)^2+\left(|\lambda_{131}|\frac{\textrm{ 100 GeV}}{m_{\tilde{\tau}_L}}\right)^2} \leq 0.66, \label{121_131}
\end{equation}
at 2$\sigma$.

Before discussing the tie-in of Eq. (\ref{121_131}) with other limits, we look next at the independent limits set by the results for $\overline{\nu}_e+e \rightarrow \overline{\nu}_e+e$ from reactor data. In this case, the electron mass-dependent terms are important and must be kept.  The cross section expression in Eq. (\ref{nue_bnd}) is modified by interchange of $g_L$ and $g_R$ for application to the $\overline{\nu}_e e$ case.

\subsubsection{$\overline{\nu}_{e}+e \rightarrow \overline{\nu}_{e}+e$}

The highest statistics experiment $\overline{\nu}_{e}+e \rightarrow \overline{\nu}_{e}+e$ is still that of Reines, Gurr and Sobel \cite{irvine}.  The results are presented as dimensionless factors times the SM charged current, $V-A$ expression $\sigma_{V-A}$, for the cross section for each of two kinetic energy bins. The cross section for a given recoil electron energy range is evidently the result of folding the differential cross section with respect to electron energy with the $\overline{\nu}_e$ flux \cite{Avignone:1971ub}, integrating over neutrino energies and then integrating over the recoil kinetic energy.  The experimental cross sections reported are $\sigma_{\textrm{exp}}=(0.87\pm0.25)\sigma_{V-A}$ ($1.5 \textrm{ MeV}\leq T\leq3.0\textrm{ MeV})$ and $\sigma_{\textrm{exp}}=(1.70 \pm 0.44)\sigma_{V-A}$ $(3.0\textrm{ MeV}\leq T\leq4.5\textrm{ MeV})$. The quantity thus calculated is a function of the $\vio{}$ parameters, which enter through the coupling $g_R^{\nu_e e}$, Eq. (\ref{gLgR}). 

The theoretical expressions for electron-neutrino and antineutrino scattering involve the same $\vio{}$ couplings. Thus, in the spirit of our multi-parameter, multi-experiment approach, we can combine data from LSND \cite{lsnd} and Irvine \cite{irvine} results for both $\Delta T$ bins in a way similar to what we did for $R_{\tau\mu}$ and $R_{\tau}$ in Section 3.1. The resulting constraint at the 2$\sigma$ level is:
\begin{equation}
\sqrt{\left(|\lambda_{121}|\frac{\textrm{100 GeV}}{m_{\tilde{\mu}_L}}\right)^2+\left(|\lambda_{131}|\frac{\textrm{100 GeV}}{m_{\tilde{\tau}_L}}\right)^2} \leq 0.38. \label{121_131_2}
\end{equation}
At first glance it is surprising that the $\overline{\nu}_e$ data, with larger uncertainty, produces tighter constraints than the $\nu_e$ data.  The source of the added resolving power is the $g_L g_R$ term in the cross section expressions, which plays a significant role in the low energy analysis and increases the sensitivity to the variation with respect to the $\vio{}$ parameters.

Though the bound Eq. (\ref{121_131_2}) is consistent with zero at 2$\sigma$, at the 1$\sigma$ level, given current values for $g_L$ and $g_R$, it is not.  This in itself is not of special significance, but it affords the opportunity to illustrate added implications when ``new $\vio{}$-physics'' is needed to fill a gap between SM and experiment. The joint bound from LSND and Irvine at 1$\sigma$ yields:
\begin{equation}
0.14 \leq \sqrt{\left(|\lambda_{121}|\frac{\textrm{100 GeV}}{m_{\tilde{\mu}_L}}\right)^2+\left(|\lambda_{131}|\frac{\textrm{100 GeV}}{m_{\tilde{\tau}_L}}\right)^2} \leq 0.34. \label{121_131_4}
\end{equation}
Projected onto each parameter one finds, 
\begin{eqnarray}
0.20 \leq |\lambda_{121}|\frac{\textrm{100 GeV}}{m_{\tilde{\mu}_L}}\leq 0.32& &\left(0.11 \leq |\lambda_{121}|\frac{\textrm{100 GeV}}{m_{\tilde{\mu}_L}}\leq 0.35\right), \label{proj}
\end{eqnarray}
at 1$\sigma$ (1.65$\sigma$), and similarly for $\lambda_{131}(\tilde{\tau}_{L})$.

Now we have the interesting situation that, taking the bound Eq. (\ref{121_131_4}) at face value, we can ask under what conditions are they consistent with stringent bounds on related parameters, coming from more recent data.   Examples we have already studied are the bound $|\lambda_{121}|(\textrm{100 GeV})/m_{\tilde{e}_{R}}\leq0.037$, Eq. (\ref{lam12kMSbar2}), and the bound $ |\lambda_{131}|(\textrm{100 GeV})/m_{\tilde{e}_{R}}\leq 0.071$, Eq. (\ref{l13kLepMS}).  We assumed here a variant of the SCD assumption, corresponding to the largest possible range for  individual parameters, $|\lambda_{12k}|\rightarrow |\lambda_{121}|$ and $|\lambda_{13k}|\rightarrow |\lambda_{131}|$. 

From Eq. (\ref{121_131_4}), scaling by $m_{\tilde{e}_{R}}$, we can now map out regions of ($m_{\tilde{\mu}_{L}}, m_{\tilde{\tau}_{L}}$) space where the constraints from Eqs. (\ref{121_131_4}), (\ref{lam12kMSbar2}) and (\ref{l13kLepMS}) are all satisfied, ignoring for present purposes the mixing of 1$\sigma$ and 2$\sigma$ constraints.  The result of this analysis is displayed in Fig. 4. To satisfy the lower bound shown in Eq. (\ref{121_131_4}), we see that at least one of the masses must be less than the mass $m_{\tilde{e}_{R}}$\footnote{This mass pattern, if it were borne out by future experimental constraints, would contradict models that predict $m_{\tilde{e}_{R}}<m_{\tilde{e}_{L}}$, as is the case in minimal supergravity, for example \cite{mSUGRA}.}. The smaller the mass becomes, the larger the other must be to satisfy the inequalities. This mass information can {\it only} be obtained if the strict SCD approach is relaxed, as we have done here. 
\begin{figure}[h!]
 \begin{center}
  \includegraphics[width=80mm, height=70mm]{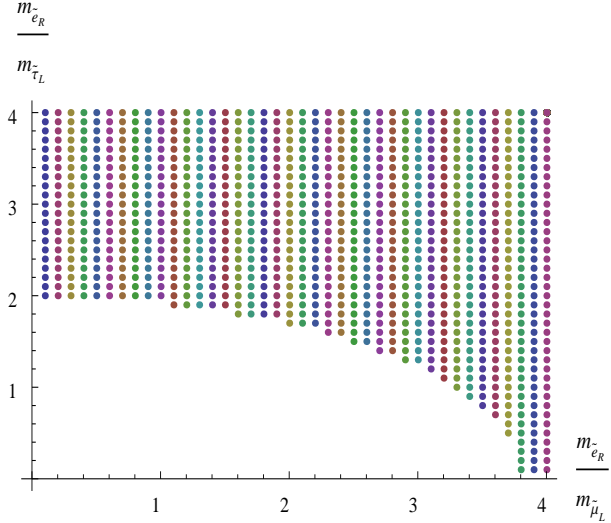}
     \caption{The darkened region shows the values of $m_{\tilde{e}_{R}}/m_{\tilde{\mu}_{L}}$ and $m_{\tilde{e}_{R}}/m_{\tilde{\tau}_{L}}$ allowed by simultaneous application of bounds shown in Eqs. (\ref{121_131_4}), (\ref{lam12kMSbar2}) and (\ref{l13kLepMS}).}
   \end{center}
\end{figure}

Alternatively, taking the bounds one at a time, Eq. (\ref{proj}), we find $m_{\tilde{\mu}_{L}}\leq 0.185$ $(0.34)$ $m_{\tilde{e}_{R}}$ or $m_{\tilde{\tau}_{L}}\leq 0.355$ $(0.65)$ $m_{\tilde{e}_{R}}$ at 1$\sigma$ (1.65$\sigma$).

The preceding discussion, summarized in Fig. 4, is offered to illustrate the added power that multi-parameter analysis provides to probe $\vio{}$ parameters.  Experiments delivering data with high statistics at energies of a MeV or so to study $\overline{\nu}_e e$ scattering would sharpen the picture, clarifying the possible role of $\vio{}$ SUSY in this sector of neutrino physics.  Here we are considering only low energy processes, where the four-fermion effective interactions apply, but at high energies the non-local effects of the exchanged particle must be included, directly probing the sfermion masses. This possibility is afforded by $e^+ e^- \rightarrow \nu \overline{\nu} \gamma$ results from LEP \cite{lep} and, in the future, possibly 100 GeV range ${\nu}_{\mu} e \rightarrow \nu_{\mu} e$ and $\nu_{\mu} e \rightarrow \nu_e \mu$ scattering experiments such as those proposed by NuSOnG \cite{Adams:2008cm}.   

This concludes our exploration of the multi-parameter effects in purely leptonic processes.  Next we consider some important constraints from semi-leptonic physics.

\section{Semi-leptonic case}

When R-parity violating interactions are taken into account, charge current and neutral current interaction generally involve more than one coupling at a time, and in some cases these couplings can be large and cancel each other. The lesson we take from the leptonic case is that such degeneracies can be removed by considering a subset of experiments characterized by the same R-parity couplings. Then one bounds the couplings by considering the experimental uncertainties on this subset altogether. This is even more evident when we analyze processes that involve the semi-leptonic couplings $\lambda'_{ijk}$ of Eq. (\ref{eff_Lagr_sl}). Contrary to the leptonic and hadronic cases, the couplings $\lambda'_{ijk}$ are not required by gauge invariance to have any symmetry in their indices. As a consequence, the number of effective couplings entering the Lagrangian is much greater than those appearing in Eq. (\ref{eff_Lagr_l}), as we have mentioned in Section 2. There are thus more processes that must be used simultaneously to bound the couplings. What this also means is that, due to the amazing overall accuracy of the SM predictions and the great number of tests, there are many more ways to cut down the allowed regions of parameter space. As we will see in the following standard examples, when the availability of experiments from which we can draw bounds on a particular coupling increases, the bound on the coupling tends to approach the one obtained under the SCD.

\subsection{Universality in pion and tau decay}

In the cases of semi-leptonic couplings, we can obtain
behavior similar in nature to the one depicted in Fig. 2a. The
ratio:
\begin{equation}
R_{\tau\pi}=\frac{\Gamma(\tau^{-}\rightarrow\pi^{-}\nu_{\tau})}{\Gamma(\pi^{-}\rightarrow\mu^{-}\bar{\nu}_{\mu})}=
R_{\tau\pi}^{SM}\frac{|V_{ud}+r'_{31k}(\tilde{d}_{Rk})|^{2}}{|V_{ud}+r'_{21k}(\tilde{d}_{Rk})|^{2}}\label{Rtaupi}
\end{equation}
would give in the SCD the $2\sigma$ bounds $|\lambda'_{31k}|\leq0.092$
$(\tilde{d}_{Rk})$ and $|\lambda'_{21k}|\leq0.032$
$(\tilde{d}_{Rk})$. Here, again, the uncertainty on the $\tau$ lifetime is comparable in magnitude to the one on the branching fraction to pions, and has to be taken into account. As in the leptonic case, the simultaneous presence of both couplings introduces a two-fold degeneracy. Such degeneracy can be
removed by considering the ratio \cite{Barger:1989rk}
\begin{equation}
    R_{\pi}=\frac{\Gamma(\pi^{-}\rightarrow e^{-}\bar{\nu}_{e})}{\Gamma(\pi^{-}\rightarrow\mu^{-}\bar{\nu}_{\mu})}=R_{\pi}^{SM}\left\{1+\frac{2}{V_{ud}}\left[r'_{11k}(\tilde{d}_{Rk})
    -r'_{21k}(\tilde{d}_{Rk})\right]\right\}.\label{Rpi}
\end{equation}
For the purpose of illustrating our multidimensional approach, it is convenient in this case to follow the restriction mentioned in \cite{Barger:1989rk}, 
so we use Eq. (\ref{Rpi}) to effectively place two alternative 2$\sigma$ bounds\footnote{Note here that Eq. (\ref{lam'11k}) implies a sum over couplings and exchanged particles. Neutrinoless double-beta decay places a strong independent bound on $\lambda'_{111}$, which applies to $\tilde{d}_{R}$- and $\tilde{u}_{L}$-exchange equally \cite{Mohapatra:1986su}.}
: either
\begin{equation}
    |\lambda'_{11k}|\leq0.051\left(\frac{m_{\tilde{d}_{Rk}}}{100\textrm{ GeV}}\right),\label{lam'11k}
\end{equation}
or
\begin{equation}
    |\lambda'_{21k}|\leq0.040\left(\frac{m_{\tilde{d}_{Rk}}}{100\textrm{ GeV}}\right).\label{lam'21k}
\end{equation}
As we have done in Sec. 3.1, we can combine Eqs. (\ref{Rtaupi}) and (\ref{Rpi}) to place 2$\sigma$ bounds on the region spanned by $\lambda'_{31k}$ and $\lambda'_{21k}$. The result is shown in Fig. 5a.
\begin{figure}[h!]
 \begin{center}
  \includegraphics[width=160mm, height=60mm]{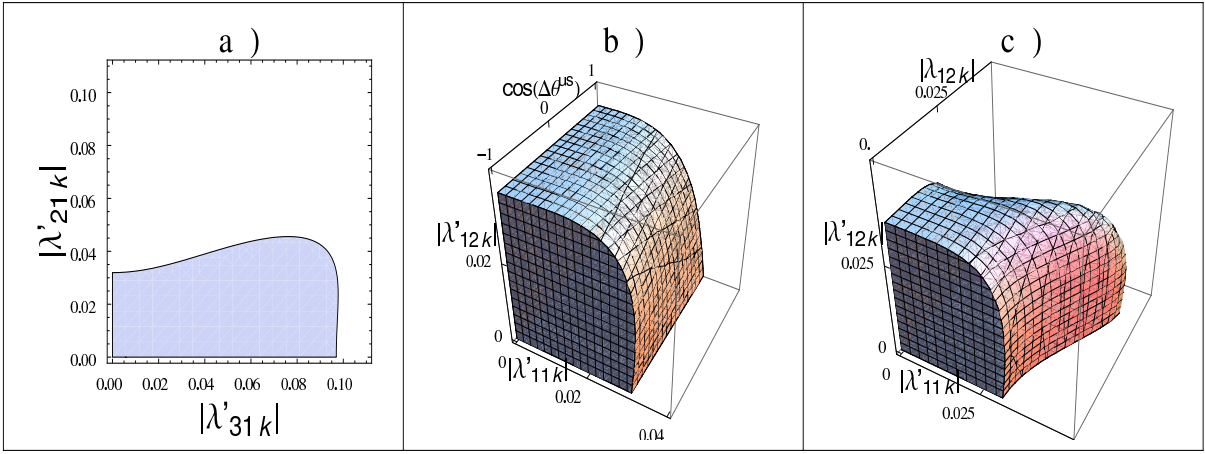}
     \caption{a) 2$\sigma$ bound region on $\lambda'_{31k}(\tilde{d}_{Rk})$, $\lambda'_{21k}(\tilde{d}_{Rk})$ from $R_{\tau\pi}$ and $R_{\pi}$ combined data, Eqs. (\ref{Rtaupi}-\ref{Rpi}). The allowed region can be enclosed in a box of size $\{|\lambda'_{31k}|(\tilde{d}_{Rk}), |\lambda'_{21k}|(\tilde{d}_{Rk})\}\leq\{0.098,0.045\}$. b) 2$\sigma$ bound region on $\lambda'_{11k}(\tilde{d}_{Rk})$, $\cos(\Delta\theta^{us}_{k})$, $\lambda'_{12k}(\tilde{d}_{Rk})$ from $R_{\pi}$, CKM unitarity (Eq. (\ref{CKM_Unit2}), $\lambda_{12k}=0$) and FB asymmetry combined data. c) 2$\sigma$ bound region on $\lambda'_{11k}$ $(\tilde{d}_{Rk})$, $\lambda_{12k}$ $(\tilde{e}_{Rk})$ and $\lambda'_{12k}$ $(\tilde{d}_{Rk})$. $|\lambda'_{12k}|$ is bounded by the FB asymmetry while $|\lambda_{12k}|$ by $\mu$ decay in the $\overline{MS}$ scheme. $\cos(\Delta\theta_{k}^{us})=-1$. The allowed region can be enclosed in a box of size $\{|\lambda'_{11k}|(\tilde{d}_{Rk}), |\lambda_{12k}|(\tilde{e}_{Rk}),|\lambda'_{12k}|(\tilde{d}_{Rk})\}\leq\{0.047,0.042,0.036\}$.}
   \end{center}
\end{figure}
The allowed region, rescaled to the masses of the exchanged squarks, can be enclosed in a box of size $\{|\lambda'_{31k}|(\tilde{d}_{Rk}), |\lambda'_{21k}|(\tilde{d}_{Rk})\}\leq\{0.098,0.045\}$. The resulting 2$\sigma$ bound on $\lambda'_{31k}$ reads:
\begin{equation}
    |\lambda'_{31k}|\leq0.092\left(\frac{m_{\tilde{d}_{Rk}}}{100\textrm{ GeV}}\right),\label{lam'31k}
\end{equation}
exactly equal to the one obtained by SCD.

\subsection{Unitarity of the CKM matrix and forward-backward
asymmetry}

The Cabibbo-Kobayashi-Maskawa (CKM) matrix elements are experimentally determined by comparing
the rates of decays that involve quarks in the initial state to
the rate of muon decay. In general, nuclear beta decay is used to
determine the value of $|V_{ud}|$, while the rates for $s\rightarrow
ul\bar{\nu}_{l}$ and $b\rightarrow ul\bar{\nu}_{l}$ in $K$ and
charmless $B$ decay are used to determine $|V_{us}|$ and $|V_{ub}|$.
The R-breaking processes involved in these decays are shown in Figure 6.

\begin{figure}[h!]
 \begin{center}
  \includegraphics[width=100mm, height=60mm]{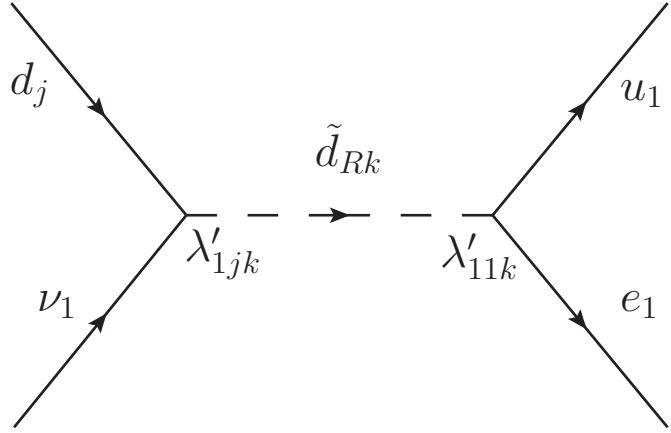}
     \caption{SUSY processes involved in $d\rightarrow
ue\bar{\nu}_{e}$ ($j=1$), $s\rightarrow
ue\bar{\nu}_{e}$ ($j=2$), and $b\rightarrow ue\bar{\nu}_{e}$ ($j=3$).}
   \end{center}
\end{figure}

The unitarity constraint can be imposed on the CKM matrix elements, together with the effective Lagrangian of Eq. 
(\ref{eff_Lagr_sl}) and a similar one, constructed from Eq. (\ref{Dirac_Lagr}), involving the product of different couplings. One gets \cite{Barbier:2004ez}:


\begin{eqnarray}
    \sum_{i=1}^{3}|V_{ud_{i}}|^2&=&\frac{1}{\left|1+r_{12k}(\tilde{e}_{Rk})\right|^{2}}
   \left(\left|V_{ud}+r'_{11k}(\tilde{d}_{Rk})\right|^{2}+\left|V_{us}+\sum_{k}\frac{\lambda'^{\ast}_{11k}\lambda'_{12k}}{4\sqrt{2}G_{F}m_{\tilde{d}_{Rk}}^{2}}\right|^{2}\right.\nonumber\\
    & &+\left.\left|V_{ub}+\sum_{k}\frac{\lambda'^{\ast}_{11k}\lambda'_{13k}}{4\sqrt{2}G_{F}m_{\tilde{d}_{Rk}}^{2}}\right|^{2}
    \right),\label{CKM_Unit1}
\end{eqnarray}
which becomes at leading order in R-parity breaking,
\begin{eqnarray}
     \sum_{i=1}^{3}|V_{ud_{i}}|^2&=&1-2r_{12k}(\tilde{e}_{Rk})+2r'_{11k}(\tilde{d}_{Rk})\left|V_{ud}\right|+2\left(\sum_{k}\frac{|\lambda'_{11k}||\lambda'_{12k}|\cos(\Delta\theta^{us}_{k})}{4\sqrt{2}G_{F}m^{2}_{\tilde{d}_{Rk}}}\right)\left|V_{us}\right|\nonumber\\
     & &+2\left(\sum_{k}\frac{|\lambda'_{11k}||\lambda'_{13k}|\cos(\Delta\theta^{ub}_{k})}{4\sqrt{2}G_{F}m^{2}_{\tilde{d}_{Rk}}}\right)\left|V_{ub}\right|,
    \label{CKM_Unit2}
\end{eqnarray}
where $\cos(\Delta\theta^{us}_{k})\equiv\cos(\theta_{us}+\theta_{12k}-\theta_{11k})$
and $\cos(\Delta\theta^{ub}_{k})\equiv\cos(\theta_{ub}+\theta_{13k}-\theta_{11k})$
are the relative phases between the CKM matrix elements and the complex R-parity violating couplings. Using Eq. (\ref{CKM_Unit2}) we can place bounds on the $\lambda'$ couplings involved by separation between the right- and left-hand side. One can substitute the most recent experimental determination of the central values of the CKM matrix element on the right, and use the errors on the unitarity bound on the left at the desired level of precision. 

In the literature, Eq. (\ref{CKM_Unit2}) is treated in the SCD,
with the additional constraint that R-parity couplings and CKM
matrix elements are treated as real. With these assumptions we find that the most recent data \cite{Amsler:2008zz} result in 
the following bounds at 2$\sigma$: $|\lambda'_{11k}|\leq0.027$
$(\tilde{d}_{Rk})$ and $|\lambda_{12k}|\leq0.028$
$(\tilde{e}_{Rk})$.

In Eqs. (\ref{CKM_Unit1}) and (\ref{CKM_Unit2}) the notation of Eq. (\ref{rijk}), to express the sum of moduli squared, has been used, together with the correction to $G_{F}$ from the muon lifetime, Eq. (\ref{GF_corr}). As can be seen, the full dependence on the CKM and R-parity violating \textit{phases} is also indicated. We have adopted the Wolfenstein parametrization \cite{Wolfenstein:1983yz} to express the CKM matrix elements. In this parametrization $V_{us}$ is real, while $V_{ub}$ is not. Nonetheless, measurements of the absolute values of the CKM elements give $|V_{ub}|\sim0.004$, approximately two orders of magnitude smaller than $|V_{ud}|\sim0.974$ and $|V_{us}|\sim0.226$. Thus, the behavior of Eq. (\ref{CKM_Unit2}) is almost independent of $\lambda'_{13k}$, as $|V_{ub}|$ can be neglected.
Taking into account the fact that $V_{us}$ is real and $|V_{ub}|$ is tiny, and neglecting for the moment the SUSY correction to $G_{F}$, Eq. (\ref{CKM_Unit2}) implies 2$\sigma$ bounds on a three dimensional parameter space spanned by $|\lambda'_{11k}|$, $|\lambda'_{12k}|$ and $\cos(\theta_{12k}-\theta_{11k})$. $\lambda'_{11k}$, can be bounded by $\pi$-decay, Eq. (\ref{Rpi}). $\lambda'_{12k}$ can be bounded by the forward-backward (FB) asymmetry in fermion pair production reactions $e^{-}e^{+}\rightarrow f\bar{f}$, which we treat in detail in the next subsection. The 2$\sigma$-bound region is shown in Fig. 5b. Note that, contrary to the other cases in the paper, here the index $k$ has to be common to the three axes in the picture. One can see that, in spite of the fact that the phases are allowed to take on any values, the $\lambda'$ parameters are allowed a slightly larger region when $\cos(\Delta\theta^{us}_{k})=-1$. We come back to this point at the end of this section. 

\subsubsection{Forward backward asymmetry}

The forward-backward asymmetry in fermion pair production has been studied at PEP, PETRA, TRISTAN, LEP, and SLC. In order to bound $\lambda'_{12k}(\tilde{d}_{Rk})$ we need charm production, $e^{-}e^{+}\rightarrow c\bar{c}$. The SUSY diagram that contributes to this process is depicted in Figure 7a, with $u_{1}\rightarrow u_{2}$, $\lambda'_{11k}\rightarrow\lambda'_{12k}$. We assume that the right-handed down squark mass is far enough above the Z-pole that we can retain our effective Lagrangians, Eqs. (\ref{eff_Lagr_sl}) and (\ref{eff_Lagr_l}), and use the data in \cite{Amsler:2008zz}, dominated by Z-pole measurements.

The SM expression for the charm FB asymmetry reads \cite{Halzen:1984mc}:
\begin{equation}
 A^{SM}_{FB}=\frac{A_{1}}{\frac{8}{3}A_{0}},\label{Afb}
\end{equation}
where
\begin{eqnarray}
 A_{0}&=&Q_{c}^{2}-\frac{Q_{c}}{2}\Re(r)\left(g^{e}_{L}g^{c}_{L}+g^{e}_{R}g^{c}_{R}+g^{e}_{L}g^{c}_{R}+g^{e}_{R}g^{c}_{L}\right)\nonumber\\
 & &+\frac{1}{4}|r|^{2}\left[(g^{e}_{L}g^{c}_{L})^{2}+(g^{e}_{R}g^{c}_{R})^{2}+(g^{e}_{L}g^{c}_{R})^{2}+(g^{e}_{R}g^{c}_{L})^{2}\right],\label{A0}\\
A_{1}&=&-Q_{c}\Re(r)\left(g^{e}_{L}g^{c}_{L}+g^{e}_{R}g^{c}_{R}-g^{e}_{L}g^{c}_{R}-g^{e}_{R}g^{c}_{L}\right)\nonumber\\
 & &+\frac{1}{2}|r|^{2}\left[(g^{e}_{L}g^{c}_{L})^{2}+(g^{e}_{R}g^{c}_{R})^{2}-(g^{e}_{L}g^{c}_{R})^{2}-(g^{e}_{R}g^{c}_{L})^{2}\right],\label{A1}
\end{eqnarray}
where $g_{L,R}$ are the usual chiral couplings, $Q_{c}=2/3$ is the charge of the charm and
\begin{equation}
 r=\frac{4\sqrt{2}G_{F}M_{Z}^{2}}{s-M^{2}_{Z}+iM_{Z}\Gamma_{Z}}\left(\frac{s}{e^{2}}\right)\label{r}
\end{equation}
parametrizes the $\gamma-Z$ interference. The R-parity contribution is obtained by the substitution:
\begin{equation}
 g^{e}_{L}g^{c}_{L}\longrightarrow g^{e}_{L}g^{c}_{L}-\frac{r'_{12k}(\tilde{d}_{Rk})}{2}.\label{r-par}
\end{equation}
So the correction to the SM reads at lowest order, 
\begin{equation}
 A_{FB}=A_{FB}^{SM}\left[1-\frac{r'_{12k}(\tilde{d}_{Rk})F(r)}{2}\left(\frac{1}{2A_{0}}-\frac{1}{A_{1}}\right)\right],\label{corr}
\end{equation}
where
\begin{equation}
 F(r)=Q_{c}\Re(r)-|r|^{2}g^{e}_{L}g^{c}_{L}\label{f-r}
\end{equation}
and $r$ has to be calculated at the Z-pole. By using the standard $SU(2)\times U(1)$ expressions for $g_{L}$ and $g_{R}$, and adopting the $\overline{MS}$ scheme value of $\sin^{2}\theta_{W}$ for definiteness, one gets the values: $g^{e}_{L}=-0.2688$, $g^{e}_{R}=0.2312$, $g^{c}_{L}=0.3459$,
$g^{c}_{R}=-0.1541$. We obtain the bound at 2$\sigma$:
\begin{equation}
    |\lambda'_{12k}|\leq 0.027\left(\frac{m_{\tilde{d}_{Rk}}}{100
    \textrm{ GeV}}\right).\label{lam'12kFBb}
\end{equation}
\smallskip

As mentioned above, Fig. 5b shows that allowing for the
$\lambda'_{11k}$ and $\lambda'_{12k}$ couplings to have opposite
complex phases ($\cos\Delta\theta_{k}^{us}=-1$) slightly extends
the allowed regions of parameter space with respect to the SCD.
Furthermore, such an extension
becomes significant when we also introduce the leptonic coupling
$\lambda_{12k}$, bounded by the experimental limits on the muon
liftime in the $\overline{MS}$ scheme, Eq. (\ref{lam12kMSbar2}). The 2$\sigma$-allowed region in $\lambda'_{11k}$, $\lambda_{12k}$ and $\lambda'_{12k}$ obtained by simoultaneous combination of the data from CKM unitarity, FB asymmetry in charm production and muon decay in the $\overline{MS}$ renormalization scheme is shown in Fig. 5c. It is enclosed in a box of size $\{|\lambda'_{11k}|(\tilde{d}_{Rk}), |\lambda_{12k}|(\tilde{e}_{Rk}),|\lambda'_{12k}|(\tilde{d}_{Rk})\}\leq\{0.047,0.042,0.036\}$, thus allowing roughly factor of two extensions of the parameters with respect to the SCD bounds. The combined analysis furnishes a new 2$\sigma$ bound on $\lambda'_{11k}$\footnote{As in Footnote 5, $\lambda'_{111}$ is tightly bounded by neutrinoless double-beta decay \cite{Mohapatra:1986su}.}:
\begin{equation}
    |\lambda'_{11k}|\leq0.039\left(\frac{m_{\tilde{d}_{Rk}}}{100
    \textrm{ GeV}}\right).\label{lam'11kCKM}
\end{equation}
This striking situation, where three parameters are all allowed to be non-zero and larger than their SCD values, is obscured when only one parameter at a time is considered, i.e. SCD is assumed uniformly.
In principle the second row of the CKM matrix could be used in a similar fashion to bound $|\lambda'_{21k}|$ and $|\lambda'_{22k}|$:
\begin{equation}
 \sum_{i=1}^{3}|V_{cd_{i}}|^2=1-2r_{12k}(\tilde{e}_{Rk})+2\left(\sum_{k}\frac{|\lambda'_{21k}||\lambda'_{22k}|\cos(\Delta\theta^{cd}_{k})}{4\sqrt{2}G_{F}m^{2}_{\tilde{d}_{Rk}}}\right)\left|V_{cd}\right|+2\sum_{i}r'_{22k}(\tilde{d}_{Rk})\left|V_{cs}\right|,\label{CKM_Urow2}
\end{equation}
where $\cos(\Delta\theta^{cd}_{k})\sim\cos(\theta_{21k}-\theta_{22k})$ in the Wolfenstein parametrization.
$|\lambda'_{21k}|$ $(\tilde{d}_{Rk})$ can be bounded by pion decay, Eq. (\ref{Rpi}).
The dependence on $\lambda_{12k}$ $(\tilde{e}_{Rk})$ comes from the bounds on universality of the Fermi constant in muon decay. We use, again, the $\overline{MS}$ bound at $2\sigma$, Eq. (\ref{lam12kMSbar2}). The weakest bound consistent with both these constraints is obtained when $\cos(\Delta\theta^{cd}_{k})=-1$ and reads, at 2$\sigma$, 
\begin{equation}
    |\lambda'_{22k}|\leq0.14\left(\frac{m_{\tilde{d}_{Rk}}}{100\textrm{ GeV}}\right).\label{lam'22kCKMCorr}
\end{equation}

A caveat is necessary at this point, in the sense that Eq. (\ref{CKM_Urow2}) is derived for processes involving the production of charmed particles in deep inelastic $\nu_{\mu}$-nucleon scattering, with the assumption of lepton flavor conservation. This is the standard textbook process used for the determination of the CKM couplings $|V_{cd}|$ and $|V_{cs}|$ \cite{Paschos:2007pi}. Such a choice is reflected by the $i=2$ index of the $\lambda'_{ijk}$ couplings entering Eq. (\ref{CKM_Urow2}). The use of recent PDG2008 data for the uncertainty affecting the unitarity constraint and for the central values of the CKM matrix elements is not fully consistent with this idealized picture. The most recent and precise values given in \cite{Amsler:2008zz} are obtained through a weighted average of different processes, some of which involve external particles of the first or third lepton generation. It is clear that the robustness of the bound given in Eq. (\ref{lam'22kCKMCorr}) depends strongly on the amount and nature of the weighting involved. Because such detailed knowledge and extensive analysis in this regard goes beyond the purposes of this paper, we limit ourselves to presenting the bound above, recommending caution in its interpretation. As we will see in Section 4.4, $D^{0}$ decay alone places bounds on the same (sum of) couplings. We consider those bounds more robust.

Finally, Eq. (\ref{CKM_Unit2}) has the nice feature that it involves the phases of the R-breaking couplings. In general such phases are associated with CP violating effects. So we can envisage a strategy that would combine additional experiments in the CP violating sector with those that can place bounds on the moduli of $\vio{}$-couplings like the two above, so that a more thorough restriction of parameter space takes place. However we did not find in the literature \cite{Barbier:2004ez}, nor were we able to create a specific example that would help us bound the phases of the couplings involved in this case, namely the product $\lambda'^{\ast}_{11k}\lambda'_{12k}$ $(\tilde{d}_{Rk}$), in terms of CP violating processes. Some asymmetries in fermion pair production at leptonic colliders ($l^{+}l^{-}\rightarrow f_{J}\bar{f}_{J'}$) on and above the $Z$-pole \cite{Chemtob:1998uq} can be expressed in term of non trivial combinations of R-breaking phases like $\Im(\lambda'^{\ast}_{1Jk}\lambda'_{1J'k}\lambda'_{ijJ}\lambda'^{\ast}_{ijJ'})/|\lambda'^{\ast}_{i'1J}\lambda'_{i'1J'}|^{2}$, with obvious summation over dummy indices.
A detailed and comprehensive study of such processes would probably shed light on the phenomenological constraints on CP violating phases. Nonetheless, due to the great number of couplings involved, such a study would have to take into account a large number of interactions, many of which cannot be treated as ``low energy'' processes. This clearly exceeds the purposes of this paper, requiring an extensive, separate investigation.

As anticipated above, and shown in Fig 5b, we have tried to constrain the phase difference $\cos(\theta_{11k}-\theta_{12k})$ by using Eq. (\ref{CKM_Unit2}), where all the absolute values are bounded by some other experiments. We have also tried to constrain the phase $\cos(\theta_{21k}-\theta_{22k})$ with the second row, Eq. (\ref{CKM_Urow2}).  We found no handle to constrain parameters, as any possible values of the phases are allowed by CKM unitarity. 


\subsection{Atomic parity violation}

We can follow the same technique, and use the bounds on
$\lambda'_{11k}$ obtained by CKM unitarity and the bounds on $\lambda_{12k}$ obtained in the $\overline{MS}$ renormalization scheme to place bounds on $\lambda'_{1j1}$ from atomic
parity violation (APV). In the SM the $Z$-exchange between the electrons and atomic nuclei leads to parity violating transitions between particular atomic levels. This has been observed for example in the $6S\rightarrow7S$
transitions of $^{133}_{55}$Cs \cite{Wood:1997zq,Guena:2004sq}. The SM contributions are encapsulated in the weak charge $Q^{SM}_{W}$, which is defined as \cite{Amsler:2008zz}
\begin{equation}
    Q^{SM}_{W}=-2\left[\left(A+Z\right)C^{SM}_{1}(u)+\left(2A-Z\right)C^{SM}_{1}(d)\right],\label{weakCh}
\end{equation}
where $Z$ is the atomic number, $A$ the atomic mass, and the
coefficients $C_{1}(i)$ are given at tree level by
\begin{eqnarray}
    C^{SM}_{1}(u)=-\frac{1}{2}+\frac{4}{3}x_{W}&\textrm{ and }&
    C^{SM}_{1}(d)=\frac{1}{2}-\frac{2}{3}x_{W}.\label{C1udSM}
\end{eqnarray}
The corresponding experimental quantities can be expressed in terms of the SM contributions and the $\vio{}$ processes depicted in Figure 7 \cite{Barger:1989rk}:
\begin{eqnarray}
    C_{1}(u)&=&C^{SM}_{1}(u)\left[1-r_{12k}\left(\tilde{e}_{Rk}\right)\right]-r'_{11k}(\tilde{d}_{Rk})\label{C1u}\\
C_{1}(d)&=&C^{SM}_{1}(d)\left[1-r_{12k}\left(\tilde{e}_{Rk}\right)\right]+r'_{1j1}(\tilde{u}_{Lj}),\label{C1d}
\end{eqnarray}
where we have assumed the R-parity correction to the Fermi constant, Eq. (\ref{GF_corr}). 
The most recent determination of the difference $\delta Q_{W}=Q^{\textrm{exp}}_{W}-Q^{SM}_{W}$ for cesium can be found in \cite{Amsler:2008zz} and its expression in terms of $\vio{}$-couplings reads:
\begin{equation}
    \delta Q_{W}=-Q^{SM}_{W}r_{12k}(\tilde{e}_{Rk})
    +376 r'_{11k}(\tilde{d}_{Rk})
    -422 r'_{1j1}(\tilde{u}_{Lj}).\label{APV}
\end{equation}

\begin{figure}[h!]
 \begin{center}
  \includegraphics[width=160mm, height=60mm]{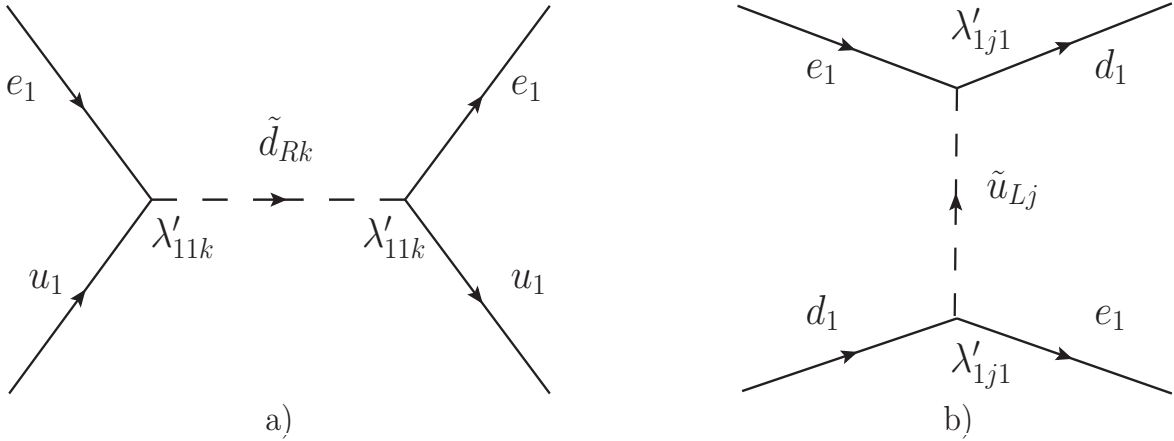}
     \caption{SUSY processes involved in Atomic Parity Violation.}
   \end{center}
\end{figure}

Again, we can first determine the $2\sigma$ bounds on the semi-leptonic couplings that one can obtain by use of the SCD: $|\lambda'_{11k}|\leq 0.051$ $(\tilde{d}_{Rk})$ and $|\lambda'_{1j1}|\leq 0.024$ $(\tilde{u}_{Lj})$. When both semi-leptonic couplings are considered, the region of parameter space that is bounded is two dimensional and its shape is similar to that of Fig. 2a. The dependence of $\delta Q_{W}$ on the leptonic coupling $\lambda_{12k}$ $(\tilde{e}_{Rk})$ due to $G_{F}$-correction introduces an additional direction in parameter space, which becomes three dimensional. $\mu$ decay in one of the renormalization schemes described in Sec. 3 can be used to place bounds on $\lambda_{12k}$, while pion decay, Eq. (\ref{Rpi}), can be used to place bounds on $\lambda'_{11k}$. The weakest bound is obtained in the $\overline{MS}$ scheme, Eq. (\ref{lam12kMSbar2}).
\begin{figure}[h!]
 \begin{center}
  \includegraphics[width=130mm, height=60mm]{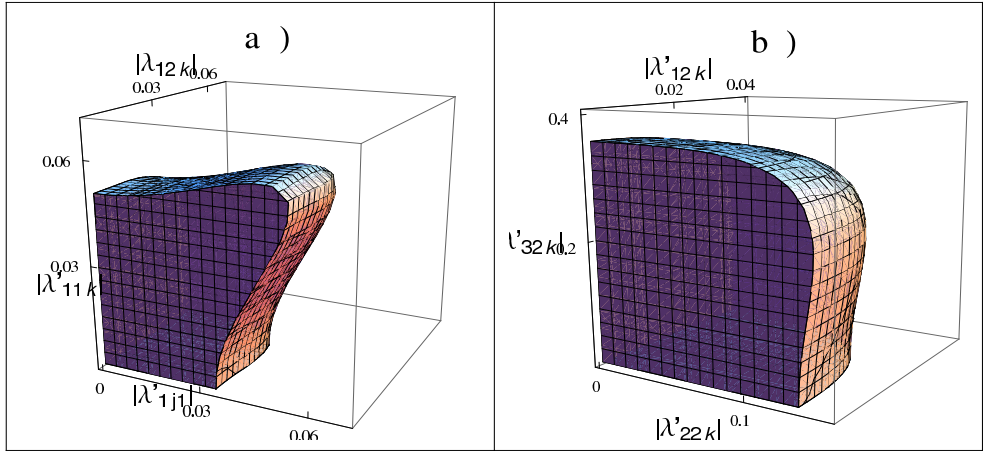}
     \caption{a) 2$\sigma$ bound region on $\lambda'_{1j1}(\tilde{u}_{Lj})$, $\lambda_{12k}(\tilde{e}_{Rk})$ and $\lambda'_{11k}(\tilde{d}_{Rk})$ from APV, $\mu$ decay in the $\overline{MS}$ renormalization scheme and $R_{\pi}$ combined. The allowed region can be enclosed in a box of size $\{|\lambda'_{1j1}|(\tilde{u}_{Lj}), |\lambda_{12k}|(\tilde{e}_{Rk}),|\lambda'_{11k}|(\tilde{d}_{Rk})\}\leq\{0.055,0.043,0.059\}$. b) 2$\sigma$ bound region on $\lambda'_{22k}(\tilde{d}_{Rk})$, $\lambda'_{12k}(\tilde{d}_{Rk})$ and $\lambda'_{32k}(\tilde{d}_{Rk})$ from $D_{0}$ decay, FB asymmetry and $D_{s}$ decay combined. The allowed region can be enclosed in a box of size $\{|\lambda'_{22k}|(\tilde{d}_{Rk}), |\lambda'_{12k}|(\tilde{d}_{Rk}),|\lambda'_{32k}|(\tilde{d}_{Rk})\}\leq\{0.140,0.034,0.359\}$.}
   \end{center}
\end{figure}
As we have explained extensively in Section 4.1, by simultaneously considering these three processes we can delimit a 2$\sigma$ bounded region of parameter space, which we present in Fig. 8a. It can be enclosed in a box of size $\{|\lambda'_{1j1}|(\tilde{u}_{Lj}), |\lambda_{12k}|(\tilde{e}_{Rk}),|\lambda'_{11k}|(\tilde{d}_{Rk})\}\leq\{0.055,0.043,0.059\}$, thus allowing only marginal extension with respect to the SCD for the $\lambda'_{11k}$ and $\lambda_{12k}$ parameters, but roughly a factor of two for the $\lambda'_{1j1}$ parameter. The 2$\sigma$ bound on $\lambda'_{1j1}$ we gather from the combined analysis reads\footnote{See Footnote 5.}:
\begin{equation}
    |\lambda'_{1j1}|\leq0.045\left(\frac{m_{\tilde{u}_{Lj}}}{100\textrm{
    GeV}}\right).\label{lambda'1j1-APV}
\end{equation}

\subsection{$\bm{D}$ decays}

For our last examples, let us now consider $D$- and $D_s$-meson decays. We can implement our procedure of taking processes that involve one or more of the couplings we have bounded in the previous cases, together with others which are at the moment unbounded, and then use the known bounds to restrict the boundaries of the allowed multidimensional parameter space to obtain bounds on the remaining couplings. 
Again we use the averages from experimental data as reported in PDG2008 and present bounds at the $2\sigma$ level. 

We consider the following ratios of
branching fractions:
$R_{D^{+}}=B(D^{+}\rightarrow\bar{K}^{0}\mu^{+}\nu_{\mu})/B(D^{+}\rightarrow\bar{K}^{0}e^{+}\nu_{e})$
and $R_{D^{0}}=B(D^{0}\rightarrow
K^{-}\mu^{+}\nu_{\mu})/B(D^{0}\rightarrow K^{-}e^{+}\nu_{e})$. 
Their expression in terms of R-breaking semileptonic couplings is given by
\begin{equation}
    \frac{R_{D^{+}}}{R_{D}^{SM}}=\frac{R_{D^{0}}}{R_{D}^{SM}}=
    \frac{\left|V_{cs}+r'_{22k}(\tilde{d}_{Rk})\right|^{2}}{\left|V_{cs}+r'_{12k}(\tilde{d}_{Rk})\right|^{2}}.\label{Ddecay}
\end{equation}
$R^{SM}_{D}=1/1.03$ is the reduction due to muon phase-space \cite{Altarelli:1997qu,Barbier:2004ez}. The bidimensional parameter space for $|\lambda'_{12k}|$ vs. $|\lambda'_{22k}|$ presents a degeneracy very similar to the one depicted in Fig. 2a, where the maximum values obtained on the axis correspond to simple use of the SCD: $|\lambda'_{22k}|\leq0.32$ $(\tilde{d}_{Rk})$, $|\lambda'_{12k}|\leq0.20$ $(\tilde{d}_{Rk})$ for $D^{+}$ decay, and $|\lambda'_{22k}|\leq0.10$ $(\tilde{d}_{Rk})$, $|\lambda'_{12k}|\leq0.21$ $(\tilde{d}_{Rk})$ for $D^{0}$ decay.


It is clear that, of the two ratios we have considered, $D^{0}$ decay places a tighter bound on these couplings. On the other hand, we can bound separately $\lambda'_{12k}$ by means of the FB asymmetry, Eq. (\ref{corr}). The bound is tight enough to cut the bidimensional degeneracy almost entirely. We get, as a consequence,
\begin{equation}
    |\lambda'_{22k}|\leq 0.090\left(\frac{m_{\tilde{d}_{Rk}}}{100
    \textrm{ GeV}}\right),\label{lam'22kDdec}
\end{equation}
ten percent stronger than that obtained by SCD.

Turning to the $D_s^{-} \rightarrow \ell^{-} +\overline{\nu}_{\ell}$ decays for further constraints,
we can bound $\lambda^{'}_{32k}$ in the same way, starting from the
ratio:
\begin{equation}
    \frac{R_{D_{s}}(\tau\mu)}{R_{D^{-}}^{SM}}=
    \frac{\Gamma(D_{s}^{-}\rightarrow\tau^{-}\bar{\nu}_{\tau})}{R_{D^{-}}^{SM}\times\Gamma(D_{s}^{-}\rightarrow\mu^{-}\bar{\nu}_{\mu})}=
    \frac{\left|V_{cs}+r'_{32k}(\tilde{d}_{Rk})\right|^{2}}{\left|V_{cs}+r'_{22k}(\tilde{d}_{Rk})\right|^{2}},\label{DsDecay}
\end{equation}
where $R_{D^{-}}^{SM}=9.76$ accounts for the phase-space suppression. One would get, by SCD use, $|\lambda'_{22k}|\leq0.27$ $(\tilde{d}_{Rk})$ and $|\lambda'_{32k}|\leq0.34$ $(\tilde{d}_{Rk})$. 
The combined analysis of $D_{0}$ decay, Eq. (\ref{Ddecay}), and Eqs. (\ref{DsDecay}) and (\ref{corr}) yields the 2$\sigma$ region depicted in Fig. 8b, whose margins are given by the box $\{|\lambda'_{22k}|(\tilde{d}_{Rk}), |\lambda'_{12k}|(\tilde{d}_{Rk}),|\lambda'_{32k}|(\tilde{d}_{Rk})\}\leq\{0.140,0.034,0.359\}$, with only slight extension beyond the bounds obtained by assuming SCD. 
This translates to the 2$\sigma$ bound: 
\begin{equation}
    |\lambda'_{32k}|\leq 0.29\left(\frac{m_{\tilde{d}_{Rk}}}{100
    \textrm{ GeV}}\right).\label{lam'32kDsdec}
\end{equation}
This bounds is, again, roughly ten percent stronger than that obtained by SCD \cite{Kundu:2008ui}. 

\section{Summary and Conclusions}

In this work, we limited our attention to experimental results from a set of standard leptonic and semi-leptonic processes and allowed $\vio{}$ parameters to vary together, constrained by data at the 2$\sigma$ level, to place bounds on their values.  We compared the resulting bounds with those obtained from the long-standing procedure of allowing only one parameter to be non-zero at a time, which has produced a long, useful list of bounds in the literature over the past twenty years or so. Using our different approach, we showed that a joint analysis of different experiments involving the same subset of couplings can explore regions of parameter space where the bounds are weakened compared to the value set by the SCD procedure. More importantly, the 2$\sigma$ bounds on individual couplings obtained by the combined approach are generally different from those obtained by strict SCD. This is due to the fact that almost all processes can by expressed in terms of more than one parameter, thus introducing correlations between the couplings and degeneracies in the allowed regions of parameter space. The combined-experiments approach helps eliminate these degeneracies and at the same time maintains the full parameter space structure. These features provide qualitatively different information from that available in the literature, whose results are almost exclusively limited to isolating parameters and considering them one at a time. New bounds obtained with our approach are given in Table 2, where we present a summary of the results described in the preceding sections.
\begin{table}
\begin{center}
\begin{tabular}{|c||c|c|c|c|}
    \hline
    $\lambda$(scale) & Experiment & Bound(2$\sigma$) & Corr. $\lambda$ & SCD Bound\\
    \hline \hline
    $\lambda_{12k}(m_{\tilde{e}_{Rk}})$ & $G_{\mu}$ & 0.037 & none & NA\\
    \hline
    $\lambda_{121}(m_{\tilde{\mu}_L})$ & $\nu_e (\overline{\nu}_e)e$ & 0.36 & $\lambda_{131}(m_{\tilde{\tau}_L})$ & 0.33\\
    \hline
    $\lambda_{121}(m_{\tilde{e}_L})$ & $\nu_{\mu}e$ & 0.118 & $\lambda_{231}(m_{\tilde{\tau}_L})$ & 0.138\\
    \hline
    $\lambda_{13k}(m_{\tilde{e}_{Rk}})$& $R_{\tau}$ & 0.071 & $\lambda_{12k}(m_{\tilde{e}_{Rk}}),\lambda_{23k}(m_{\tilde{e}_{Rk}})$ & 0.048\\
    \hline
     $\lambda_{131}(m_{\tilde{\tau}_L})$ & $\nu_e(\overline{\nu_e})e$ & 0.36 & $\lambda_{121}(m_{\tilde{\mu}_L})$ & 0.33\\
     \hline
     $\lambda_{23k}(m_{\tilde{e}_{Rk}})$ & $R_{\tau}$ & 0.066 & $\lambda_{12k}(m_{\tilde{e}_{Rk}}),\lambda_{13k}(m_{\tilde{e}_{Rk}})$ & 0.051\\
      \hline
       $\lambda_{231}(m_{\tilde{\tau}_L})$ & $\nu_{\mu}e$ & 0.118 & $\lambda_{121}(m_{\tilde{e}_{L} })$ & 0.138\\
     \hline
     $\lambda'_{11k}(m_{\tilde{d}_{Rk}})$ & $CKM_{unitary}$  & 0.039 & $\lambda_{12k}(m_{\tilde{e}_{Rk}}),\lambda'_{12k}(m_{\tilde{d}_{Rk}})$ & 0.027\\
     \hline
     $\lambda'_{12k}(m_{\tilde{d}_{Rk}})$ & $A_{FB}(c\overline{c})$  & 0.027 & none & NA\\
     \hline
     $\lambda'_{22k}(m_{\tilde{d}_{Rk}})$ & $D_0$ decay  & 0.090 & $\lambda '_{12k}(m_{\tilde{d}_{Rk}})$ & 0.10\\
     \hline
     $\lambda'_{21k}(m_{\tilde{d}_{Rk}})$ & $(\pi/\tau)_{universal.}$  & 0.040 & $\lambda '_{31k}(m_{\tilde{e}_{Rk}})$ & 0.032\\
     \hline
     $\lambda'_{31k}(m_{\tilde{d}_{Rk}})$ & $(\pi/\tau)_{universal.}$  & 0.092 & $\lambda '_{21k}(m_{\tilde{e}_{Rk}})$ & 0.092\\
     \hline
      $\lambda'_{1j1}(m_{\tilde{u}_{Lj}})$ & APV & 0.045 & $\lambda_{12k}(m_{\tilde{e}_{Rk}}),\lambda'_{11k}(m_{\tilde{d}_{Rk}})$ & 0.024\\
     \hline
      $\lambda'_{32k}(m_{\tilde{d}_{Rk}})$ & $D_s$ decay  & 0.29 & $\lambda '_{22k}(m_{\tilde{d}_{Rk}}),\lambda'_{12k}(m_{\tilde{d}_{Rk}})$ & 0.34\\
     \hline
     
\end{tabular}
\caption{Summary of constraints on $\lambda$ values with their corresponding mass scale in parenthesis. The ``Experiment'' column gives the measured quantities that are the source of the multi-variable bound, the ``Bound'' column.  The ``Corr. $\lambda$'' column gives the most directly correlated $\lambda$ determining the constraint, while the final column, ``SCD bound'' gives the value of the bound when all the relevant $\lambda$ couplings but the one in the first column are set to zero.  The note ``none'' in a column means that only one coupling appears in the relevant expression to compare to experiment.  The note ``NA'' means that there is no other coupling to set to zero for the case in this row.}

\end{center}
\end{table}

In the $\bar{\nu}_{e}e$ case, we found that the requirement that certain trilinear couplings were non-zero, combined with simultaneous constraints involving the same couplings but different sfermion masses, we could extract hierarchical relationships among these masses. We illustrated this situation in Fig. 4, where the 1$\sigma$ allowed area in the space of ``mass ratios'' is displayed, and in the paragraphs following Eq. (\ref{proj}), where individual 1$\sigma$ and 1.65$\sigma$ mass bounds are shown. To the best of our knowledge, this is the first effort, in the context of purely phenomenological bounds on $\vio{}$ parameters, to find constraints among the sfermion masses. 

In conclusion, we can say that, overall, a richer, more complex picture of parameter space and, in most cases weaker bounds on $\vio{}$ parameters result from a multi-parameter, multi-process analysis, compared to the analysis of each parameter in isolation.  This conclusion is non-trivial when, as the case in expressions we consider, parameters enter as sums of squares, suggesting that dropping all parameters but one provides the most conservative limit on each. Nonetheless, since we found the allowed ranges of parameters were larger by at most a factor two, we conclude that the SCD approach is a reliable order of magnitude estimate of the upper bounds on the individual parameters.  At the same time, we conclude that fuller analyses, as exemplified here, are needed to search for hints that data are showing R-parity violation in a region of parameter space where several parameters are non-zero. Finally, such analyses are needed to explore mass relations among sparticle masses, which requires disentangling couplings and masses by comparisons of theory with data.

{\em Acknowledgements:} We thank Danny Marfatia for many helpful
conversations on this work, and Azar Mustafayev for discussions about models of SUSY mass patterns. We acknowledge the use of the program JaxoDraw \cite{Binosi:2003yf} for drawing the Feynman diagrams of this paper. F.T. would like to thank the University of Kansas Dept. of Physics and Astronomy for hospitality while a visitor financed by the Higher Education Commission, Islamabad, Pakistan: Post-Doctoral Fellowship grant No. 1-3/PM-PDFP/2006/129. Additional support was provided from NSF Grant No. PHY-0544278 and DOE Grant No. DE-FG02-04ER41308. E.S. was also supported in part by NSF Grant No. PHY-0544278. D.W.M. received support from DOE Grant No. DE-FG02-04ER41308.

\end{document}